\documentclass[prd,preprint,tightenlines,floatfix,showpacs,preprintnumbers,nofootinbib,eqsecnum]{revtex4}

 \usepackage[dvips,final]{graphicx}
  \usepackage{amssymb}
   \usepackage{amsmath}
    \usepackage{amsfonts}
     \usepackage{epsfig}
      \usepackage{bm}

\usepackage[section]{placeins}

\usepackage{multirow}
\usepackage{ctable}
\usepackage{booktabs}
\usepackage{array}
\usepackage{tabularx}
\usepackage{xcolor}
\usepackage{pstricks}

\begin{document}

\vfill
\title{Single-parton scattering versus double-parton scattering \\
in the production of two $c \bar c$ pairs \\
and charmed meson correlations at the LHC}

\author{Andreas van Hameren}
\email{andreas.hameren@ifj.edu.pl} \affiliation{Institute of Nuclear Physics PAN, PL-31-342 Cracow, Poland}

\author{Rafa{\l} Maciu{\l}a}
\email{rafal.maciula@ifj.edu.pl} \affiliation{Institute of Nuclear Physics PAN, PL-31-342 Cracow, Poland}

\author{Antoni Szczurek}
\email{antoni.szczurek@ifj.edu.pl} \affiliation{Institute of Nuclear Physics PAN, PL-31-342 Cracow, Poland and\\
University of Rzesz\'ow, PL-35-959 Rzesz\'ow, Poland}

\date{\today}

\begin{abstract}
We compare results of exact calculations of single parton scattering (SPS) 
and double parton scattering (DPS) for production of 
$c \bar c c \bar c$ and for $D$ meson correlations. 
The SPS calculations are performed in collinear approximation
with exact matrix element for $g g \to c \bar c c \bar c$
and $q \bar q \to c \bar c c \bar c$ subprocesses.
It is shown that the contribution of gluon-gluon subprocess
is about factor 50 larger than that for quark-antiquark annihilation.
The new results are compared with results of previous calculation
with the approximate matrix element for $g g \to c \bar c c \bar c$ 
in the high-energy approximation. 
The cross section for the present exact calculation is bigger only 
at small invariant masses and small rapidity difference between
two $c$ quarks (or two $\bar c$ antiquarks).
We compare correlations in rapidities of two $c$ (or two $\bar c$) for
DPS and SPS contributions. 
Finally we compare our predictions for $D$ mesons with recent results 
of the LHCb collaboration for invariant mass, rapidity distance between
mesons and dimeson invariant mass. The predicted shapes are similar
to the measured ones, however, some strength seems to be lacking.
Our new calulations clearly confirm the dominance of DPS in the
production of events with double charm.
\end{abstract}


\maketitle

\section{Introduction}

It was advocated in Ref.~\cite{Luszczak:2011zp} that the cross section
for $c \bar c c \bar c$ production at high energy may be very large
due to double parton scattering (see also \cite{Cazaroto:2013fua}). 
The first evaluation was performed in leading order in collinear approximation.
In the meantime the LHCb collaboration measured the
cross section for the production of $D^0 D^0$ pairs at 
$\sqrt{s}$ = 7 TeV which is surprisingly large \cite{Aaij:2012dz}.
Some interesting differential studies were performed there.

Somewhat later two of us discussed several differential distributions 
for double charm production in the $k_t$-factorization approach 
using unintegrated gluon distributions \cite{Maciula:2013kd} and 
found several observables useful to identify the DPS effects. 
So far the single parton scattering contribution 
to $c \bar c c \bar c$ was calculated only in high-energy
approximation \cite{Schafer:2012tf}. For kinematical reasons the approximation 
should be reasonable for large rapidity distances. In real experiments
the condition of large rapidity distances is not always fulfilled.

The double scattering effects were studied in several other processes
such as four jet production 
\cite{DPSjets-Berger,DPSjets-Wiedemann,DPSjets-Strikman}, 
production of $W^+ W^-$ pairs
\cite{DPSgauge-Stirling,DPSgauge-Treleani}, 
production of four charged leptons
\cite{DPSleptons-Maina,DPSleptons-Stirling,DPSmuons-Stirling}. 
In all the cases the DPS contributions are much smaller than
the SPS contributions. 
The production of double hidden charm was studied e.g. in
Ref.~\cite{Baranov:2012re} for the $p p \to J/\psi J/\psi X$ process.
There the SPS and DPS contributions are comparable. The DPS contribution
exceeds the SPS contribution for large rapidity distance between
the two $J/\psi$'s. This is similar for $c \bar c c \bar c$ production
\cite{Maciula:2013kd}.

The aim of the present study is to make a detailed comparison of
the DPS and SPS with full (exact) matrix element.
The results presented here are obtained with
a code which automatically generates matrix element.
In the present study we include only $g g \to c \bar c c \bar c$ 
subprocess which is sufficient at high energies.   
We intend to make a detailed comparison of the present SPS results
to the results obtained in the high-energy approximation \cite{Schafer:2012tf}.
Finally our results will be compared to recent LHCb data \cite{Aaij:2012dz}.

\section{Some details of the calculation}

In leading-order collinear approximation the differential distributions 
for $c\bar{c}$ production depend e.g. on the rapidity of the quark, 
the rapidity of the antiquark and the transverse momentum of one of 
them (they are identical) \cite{Luszczak:2011zp}. 
In the next-to-leading order (NLO) collinear approach or in the
$k_t$-factorization approach the situation is more complicated as there 
are more kinematical variables necessary to describe the kinematical situation.
In the $k_t$-factorization approach the differential cross section for
DPS production of $c \bar c c \bar c$ system, assuming factorization of the DPSmodel, can be written as: 
\begin{eqnarray}
\frac{d \sigma^{DPS}(p p \to c \bar c c \bar c X)}{d y_1 d y_2 d^2 p_{1,t} d^2 p_{2,t} 
d y_3 d y_4 d^2 p_{3,t} d^2 p_{4,t}} = \nonumber \;\;\;\;\;\;\;\;\;\;\;\;\;\;\;\;\;\;\;\;\;\;\;\;\;\;\;\;\;\;\;\;\;\;\;\;\;\;\;\;\;\;\;\;\;\;\;\;\;\;\;\;\;\;\;\;\;\;\;\;\;\;\;\;\;\;\; \\ 
\frac{1}{2 \sigma_{eff}} \cdot
\frac{d \sigma^{SPS}(p p \to c \bar c X_1)}{d y_1 d y_2 d^2 p_{1,t} d^2 p_{2,t}}
\cdot
\frac{d \sigma^{SPS}(p p \to c \bar c X_2)}{d y_3 d y_4 d^2 p_{3,t} d^2 p_{4,t}}.
\end{eqnarray}

When integrating over kinematical variables one obtains
\begin{equation}
\sigma^{DPS}(p p \to c \bar c c \bar c X) = \frac{1}{2 \sigma_{eff}}
\sigma^{SPS}(p p \to c \bar c X_1) \cdot \sigma^{SPS}(p p \to c \bar c X_2).
\label{basic_formula}
\end{equation}
These formulae assume that the two parton subprocesses are not
correlated. 
The parameter $\sigma_{eff}$ in the denominator of above formulae can 
be defined in the impact parameter space as:
\begin{equation}
\sigma_{eff} = \left[ \int d^{2}b \; (T(\vec{b}))^{2} \right]^{-1},
\end{equation}
where the overlap function
\begin{equation}
T ( \vec{b} ) = \int f( \vec{b}_{1} ) f(\vec{b}_{1} - \vec{b} ) \; d^2 b_{1},
\end{equation}
of the impact-parameter dependent double-parton distributions (dPDFs) 
are written in the following factorized approximation 
\cite{GS2010,Gustaffson2011}:
\begin{equation}
\Gamma_{i,j}(x_1,x_2;\vec{b}_{1},\vec{b}_{2};\mu_{1}^{2},\mu_{2}^{2}) = F_{i,j}(x_1,x_2;\mu_{1}^{2},\mu_{2}^{2}) \; f(\vec{b}_{1}) \; f(\vec{b}_{2}).
\end{equation}
Then the impact-parameter distribution can be written as
\begin{equation}
\Gamma(b,x_1,x_2;\mu_1^2,\mu_2^2) = F(x_1,\mu_1^2) \; F(x_2,\mu_2^2) \;
F(b;x_1,x_2,\mu_1^2,\mu_2^2), \;
\label{correlation_function}
\end{equation}
where $b$ is the parton separation in the impact parameters space.
In the formula above the function $F(b;x_1,x_2,\mu_1^2,\mu_2^2)$
contains all information about correlations between the two partons
(two gluons in our case). The dependence was studied
numerically in Ref.~\cite{Gustaffson2011} within the Lund Dipole Cascade model.
The biggest discrepancy was found in the small $b$ region,
particularly for large $\mu_1^2$ and/or $\mu_2^2$. 
In general, the effective cross section may  
depend on kinematical variables:
\begin{equation}
\sigma_{eff}(x_1,x_2,x'_1,x'_2,\mu_1^2,\mu_2^2) =
\left(
\int d^2 b \; F(b;x_1,x_2,\mu_1^2,\mu_2^2) \; F(b;x'_1,x'_2,\mu_1^2,\mu_2^2)
\right)^{-1}.
\label{generalized_sigma_eff}
\end{equation}
The effect discussed in Ref.~\cite{Gustaffson2011} may 
give $\sim$ 10 - 20 \% on integrated cross section and even more 
$\sim$ 30 - 50 \% in some particular parts of the phase space.

In the present study we concentrate, however, rather on higher-order 
corrections and ignore the interesting dependence of the impact factors 
on kinematical variables. The dependence may be different for different
dynamical models used.

Gaunt and Stirling \cite{GS2010} also ignored the dependence of the impact
factors, but included the evolution of the double-parton distribution
amplitudes. In our previous paper \cite{Luszczak:2011zp} we have shown 
that the evolution has very small impact on the cross section 
for $p p \to c \bar c c \bar c X$.

Experimental data from Tevatron \cite{Tevatron} and LHC \cite{Aaij:2011yc,Aaij:2012dz,Aad:2013bjm} 
provide an estimate of $\sigma_{eff}$ in the denominator of formula 
(\ref{basic_formula}). A detailed analysis of $\sigma_{eff}$
based on the experimental data can be found in Ref.~\cite{Seymour:2013qka,Bahr:2013gkj}
 
In our analysis we take $\sigma_{eff}$ = 15 mb. In the most general case one 
may expect some violation of this simple factorized Ansatz given by 
Eq.~\ref{basic_formula} \cite{Gustaffson2011}.

In the present approach we concentrate on high energies and therefore 
ignore the quark induced processes.
They could be important only at extremely large pseudorapidities, very
large transverse momenta and huge $c \bar c$ invariant masses. 
In the present analysis we avoid these regions of phase space.

In our present analysis cross section for each step is calculated in the
$k_t$-factorization approach (as in Ref.~\cite{Maciula:2013wg}), that is:
\begin{eqnarray}
\frac{d \sigma^{SPS}(p p \to c \bar c X)}{d y_c d y_{\bar{c}} d^2 p_{c,t} d^2 p_{\bar{c},t}} 
&& = \frac{1}{16 \pi^2 {\hat s}^2} \int \frac{d^2 k_{1t}}{\pi} \frac{d^2 k_{2t}}{\pi} \overline{|{\cal M}_{g_{1}^{*} g_{2}^{*} \rightarrow c \bar{c}}|^2} \nonumber \\
&& \times \;\; \delta^2 \left( \vec{k}_{1t} + \vec{k}_{2t} - \vec{p}_{c,t} - \vec{p}_{\bar{c},t}
\right)
{\cal F}(x_1,k_{1t}^2,\mu^2) {\cal F}(x_2,k_{2t}^2,\mu^2),
\nonumber
\end{eqnarray}
%
%
%
The matrix elements for $g^* g^* \to c \bar c$ (off-shell gluons) 
must be calculated including transverse momenta of initial gluons 
as it was done first in Refs.~\cite{CCH91,CE91,BE01}.
The unintegrated ($k_t$-dependent) gluon distributions (UGDFs) in the
proton are taken from the literature \cite{KMR,KMS,Jung}. 
Due to the emission of extra gluons encoded in these objects, it is
belived that a sizeable part of NLO corrections is effectively
included. The framework of the $k_t$-factorization approach is 
often used with success in describing inclusive spectra of
$D$ or $B$ mesons as well as for theoretical predictions for so-called 
nonphotonic leptons, products of semileptonic decays of charm and 
bottom mesons \cite{Teryaev,Baranov00,BLZ,Shuvaev,LMS09,MSS2011,JKLZ}.

Now we go to the SPS production mechanisms of $c \bar c c \bar c$.
The elementary cross section for the SPS mechanism of double $c \bar c$
production has the following generic form:
\begin{equation}
d\hat{\sigma} = \frac{1}{2\hat{s}} \; \overline{|{\cal M}_{g g \rightarrow c \bar{c} c \bar{c}}|^2} \; d^{4} PS .
\end{equation}
where
\begin{equation}
d^{4} PS = \frac{d^3 p_1}{E_1 (2 \pi)^3} \frac{d^3 p_2}{E_2 (2 \pi)^3}
           \frac{d^3 p_3}{E_3 (2 \pi)^3} \frac{d^3 p_4}{E_4 (2 \pi)^3}
           \delta^4 \left( p_1 + p_2 + p_3 + p_4 \right) \; . 
\end{equation}
Above $p_1, p_2, p_3, p_4$ are four-momenta of final 
$c$, $\bar c$, $c$, $\bar c$ quarks and antiquarks, respectively.

Neglecting small electroweak corrections, the hadronic cross section 
is then the integral
\begin{eqnarray}
d \sigma &=& \int d x_1 d x_2 (
           g(x_1,\mu_F^2) g(x_2,\mu_F^2) 
           \; d \sigma_{gg \to c \bar c c \bar c} \nonumber \\
         &+& \Sigma_f \; q_f(x_1,\mu_F^2) \bar q_f(x_2,\mu_F^2) 
           \; d \sigma_{q \bar q \to c \bar c c \bar c} \nonumber \\
         &+& \Sigma_f \; \bar q_f(x_1,\mu_F^2) q_f(x_2,\mu_F^2)
           \; d \sigma_{\bar q q \to c \bar c c \bar c} )
      \; .
\label{hadronic cross section}
\end{eqnarray} 
In the calculation below we include 
$u \bar u$, $\bar u u$, $d \bar d$, $\bar d d$, $s \bar s$, $\bar s s$ 
annihilation terms. 
In the following we shall discuss uncertainties related to the choice 
of factorization scale $\mu_F^2$.

The matrix elements for single-parton scattering
were calculated using color-connected helicity amplitudes. They allow for an
explicit exact sum over colors, while the sum over helicities can be
dealt with using Monte Carlo methods. The color-connected amplitudes
were calculated with an automatic program similar to
{\sc Helac}~\cite{Kanaki:2000ey,Cafarella:2007pc}, following a recursive
numerical Dyson-Schwinger approach. Phase space integration was
performed with the help of {\sc Kaleu}~\cite{vanHameren:2010gg}, which
automatically generates importance sampled phase space points.

\section{Numerical Results}
\label{sec:results}

\subsection{$c \bar c c \bar c$ production}

In this subsection we discuss quark level cross sections.
It is our aim to make a detailed comparison of SPS and DPS
contributions.
Let us start from single-particle $c$ (or $\bar c$) distributions.

\begin{figure}[!h]
\begin{minipage}{0.47\textwidth}
 \centerline{\includegraphics[width=1.0\textwidth]{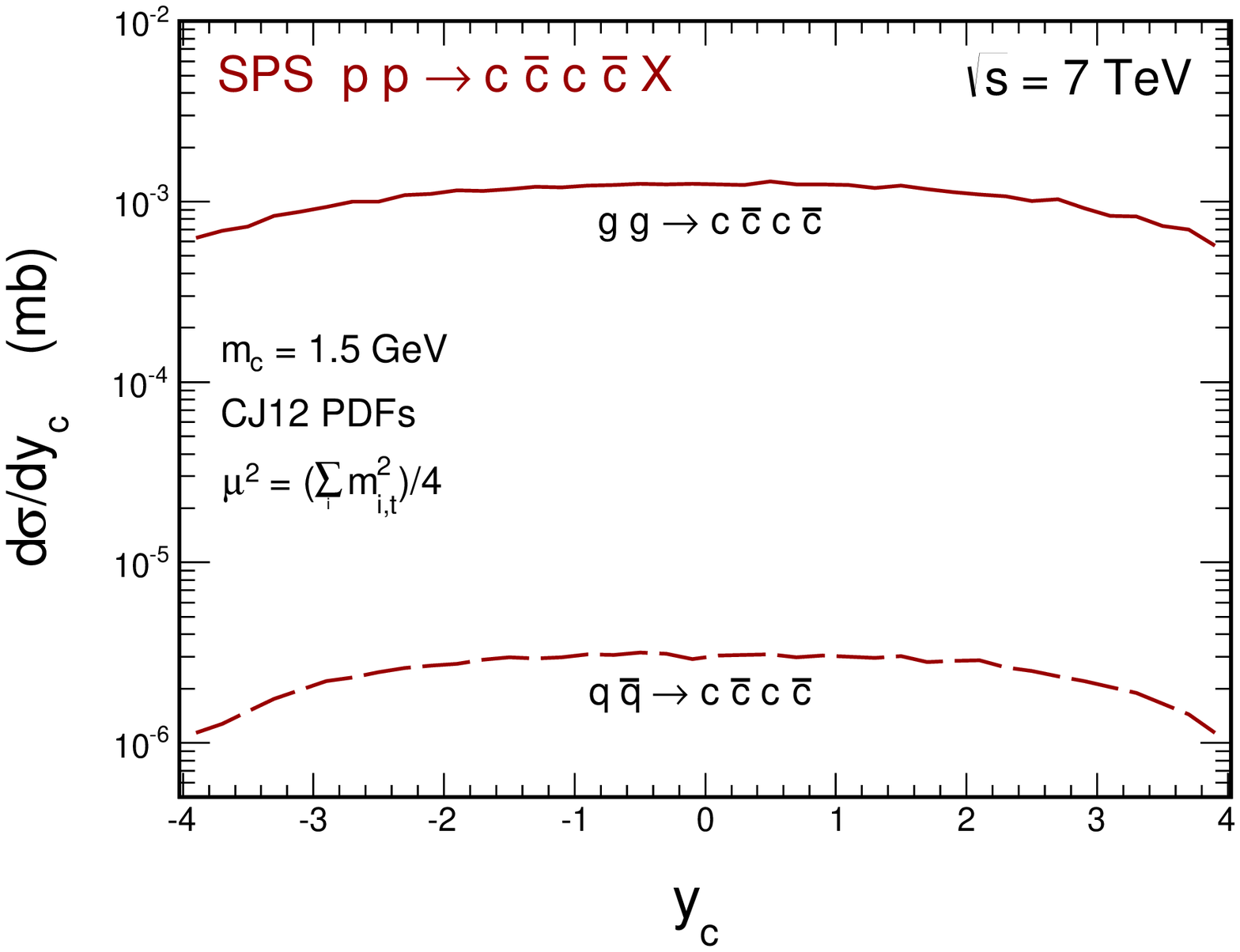}}
\end{minipage}
\hspace{0.5cm}
\begin{minipage}{0.47\textwidth}
 \centerline{\includegraphics[width=1.0\textwidth]{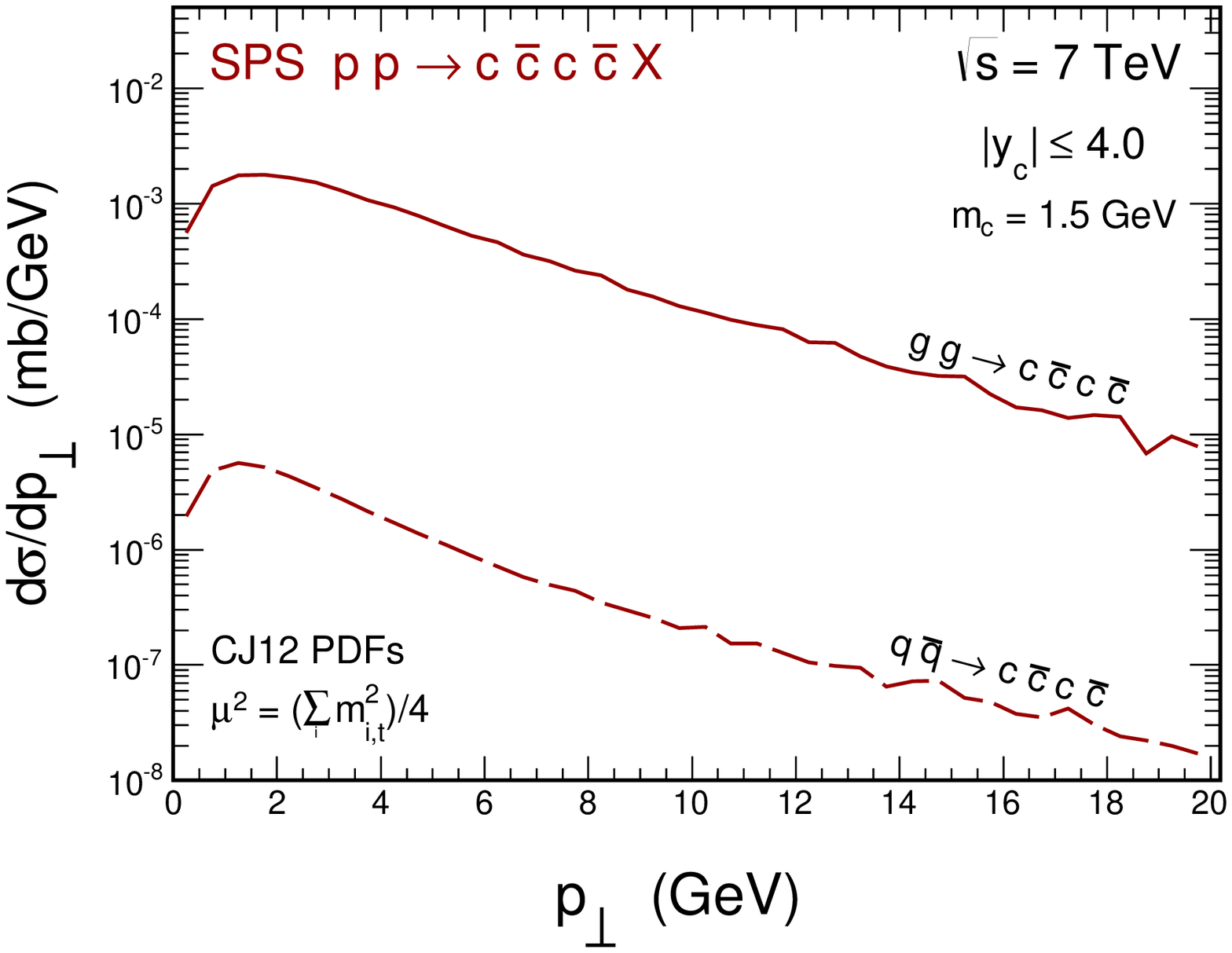}}
\end{minipage}
   \caption{
\small Rapidity (left panel) and transverse momentum (right panel)
distributions of charm quarks for $c \bar c c \bar c$ production.
We compare contributions of gluon fusion and quark-antiquark annihilation.
}
 \label{fig:pt-and-y-spectra-qqbar}
\end{figure}

In all calculations of the SPS contribution the CJ12 parton distribution
functions are used \cite{CJ12}.
In Fig.~\ref{fig:pt-and-y-spectra-qqbar} we compare contributions
for $g g \to c \bar c c \bar c$ and $q \bar q \to c \bar c c \bar c$
subproceses to distributions in rapidity and transverse momentum 
of one of $c$ quark (or $\bar c$ antiquark) at $\sqrt{s}$ = 7 TeV.
As for single $c \bar c$ production \cite{LMS09} the
gluon-gluon fusion gives the contributions of almost two-orders of 
magnitude larger than the quark-antiquark (antiquark-quark) annihilation.
The shapes are, however, rather similar.

\begin{figure}[!h]
\begin{minipage}{0.47\textwidth}
 \centerline{\includegraphics[width=1.0\textwidth]{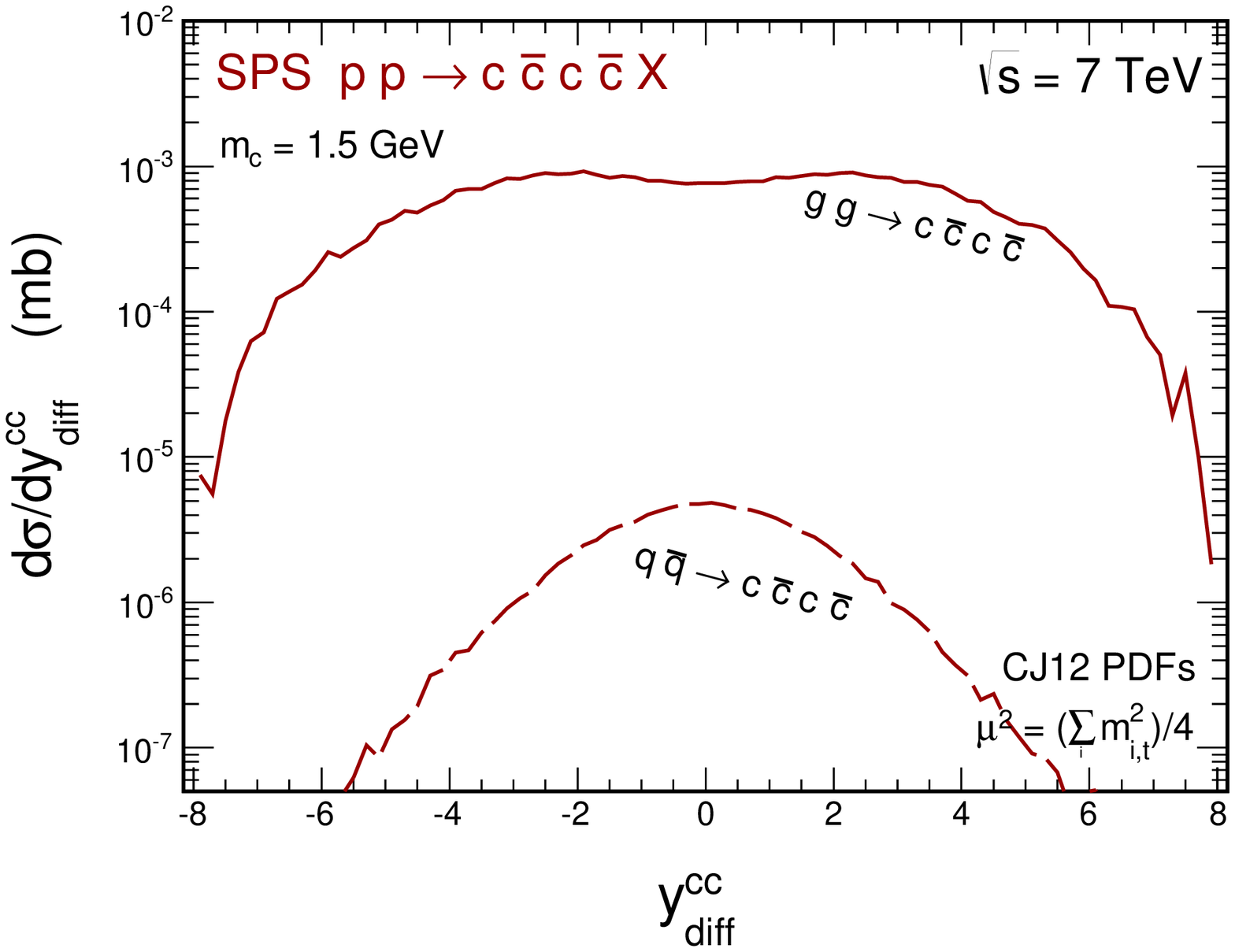}}
\end{minipage}
\hspace{0.5cm}
\begin{minipage}{0.47\textwidth}
 \centerline{\includegraphics[width=1.0\textwidth]{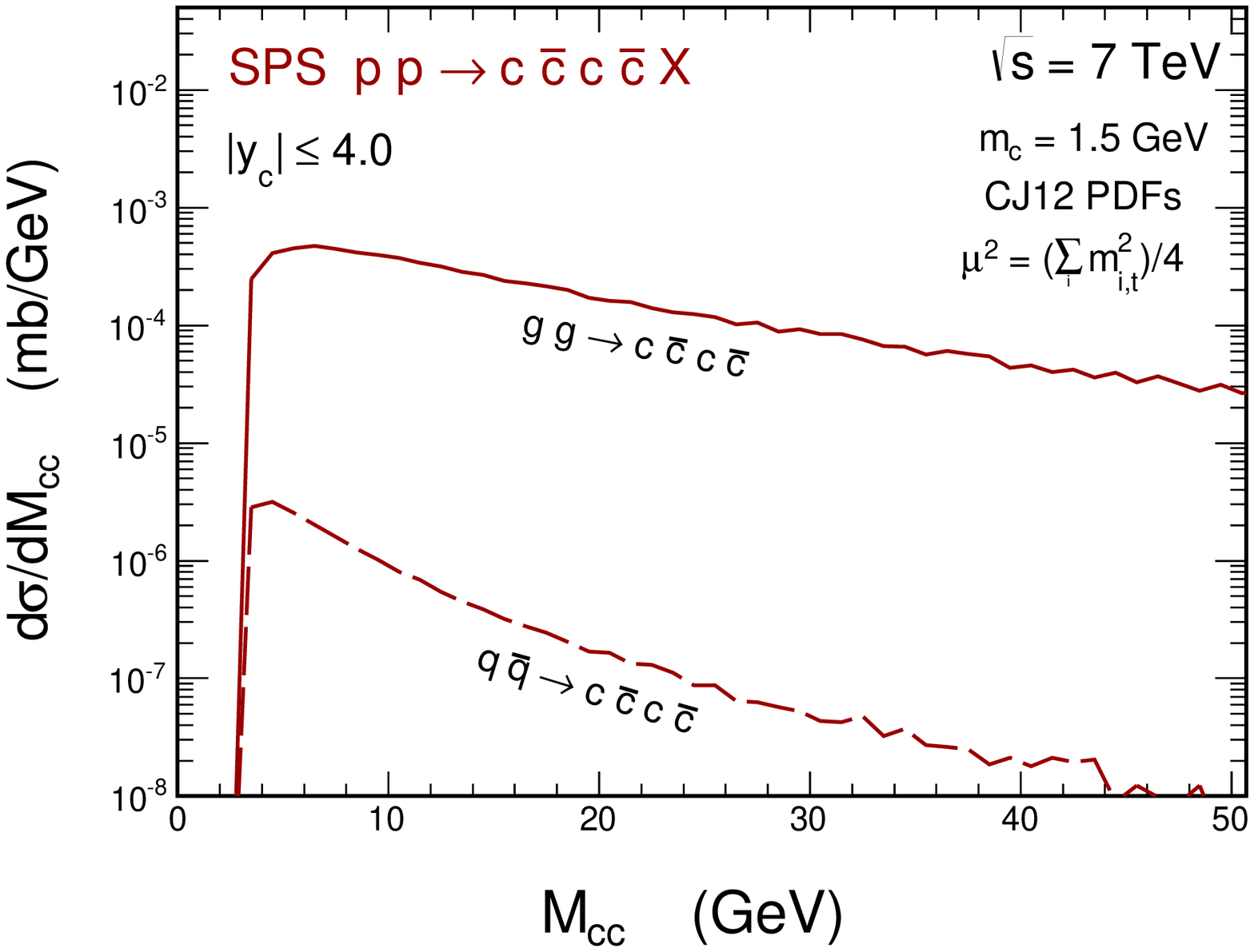}}
\end{minipage}
   \caption{
\small Distribution in rapidity distance between two $c$ quarks 
(or two $\bar c$ antiquarks) and in invariant mass of $c c$ 
(or $\bar c \bar c$) system.
}
 \label{fig:ydiff-and-minv-spectra-qqbar}
\end{figure}

Fig.~\ref{fig:ydiff-and-minv-spectra-qqbar} shows distributions in
rapidity distance between two $c$ quarks (or two $\bar c$ antiquarks)
and distribution in $c c$ (or $\bar c \bar c$) invariant mass.
The quark-antiquark component is concentrated at smaller rapidity
distances and/or smaller invariant masses than the gluon-gluon
component. However, in practice the quark-antiquark terms are at
the LHC energies negligible and will be ignored in the following.

\begin{figure}[!h]
\begin{minipage}{0.47\textwidth}
 \centerline{\includegraphics[width=1.0\textwidth]{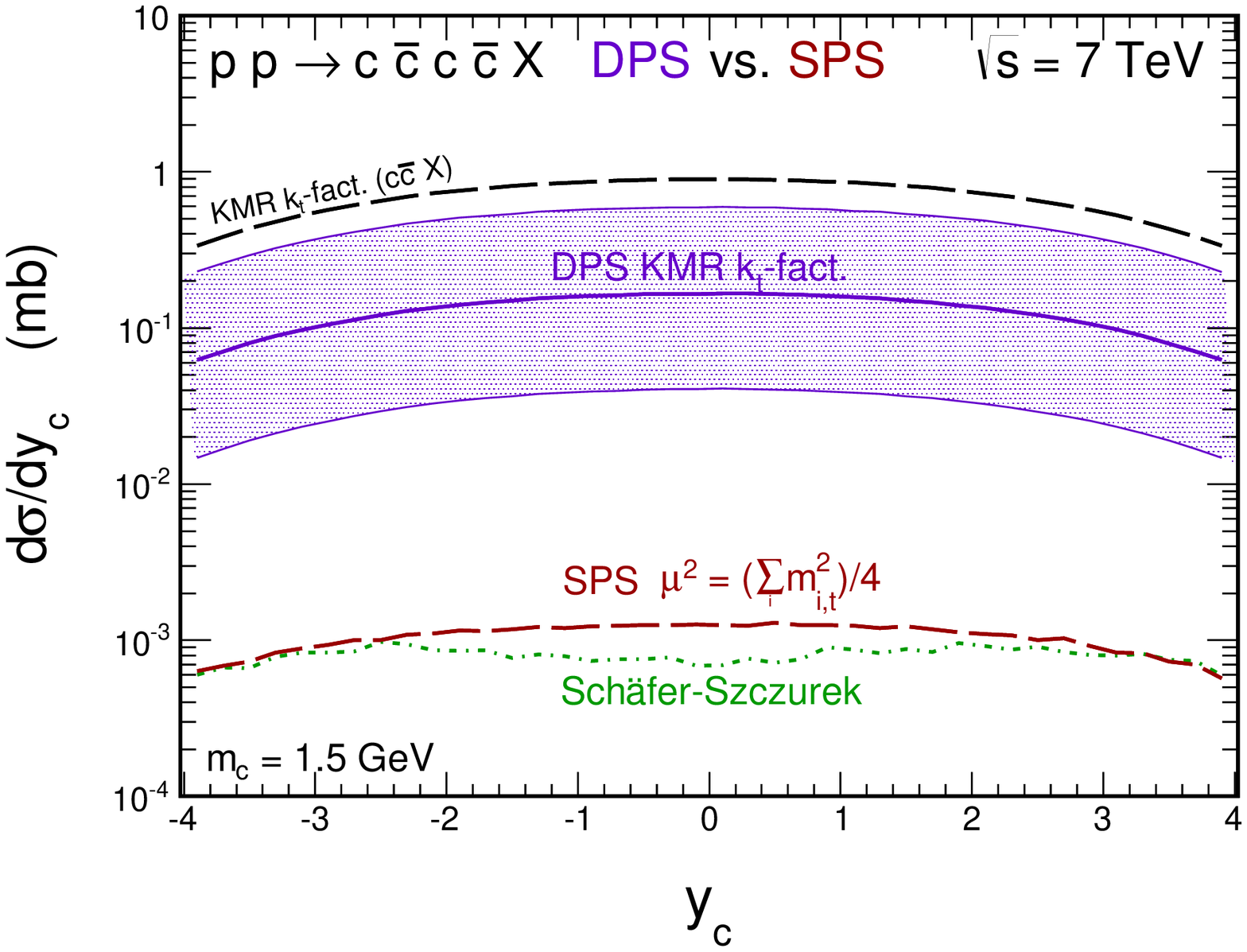}}
\end{minipage}
\hspace{0.5cm}
\begin{minipage}{0.47\textwidth}
 \centerline{\includegraphics[width=1.0\textwidth]{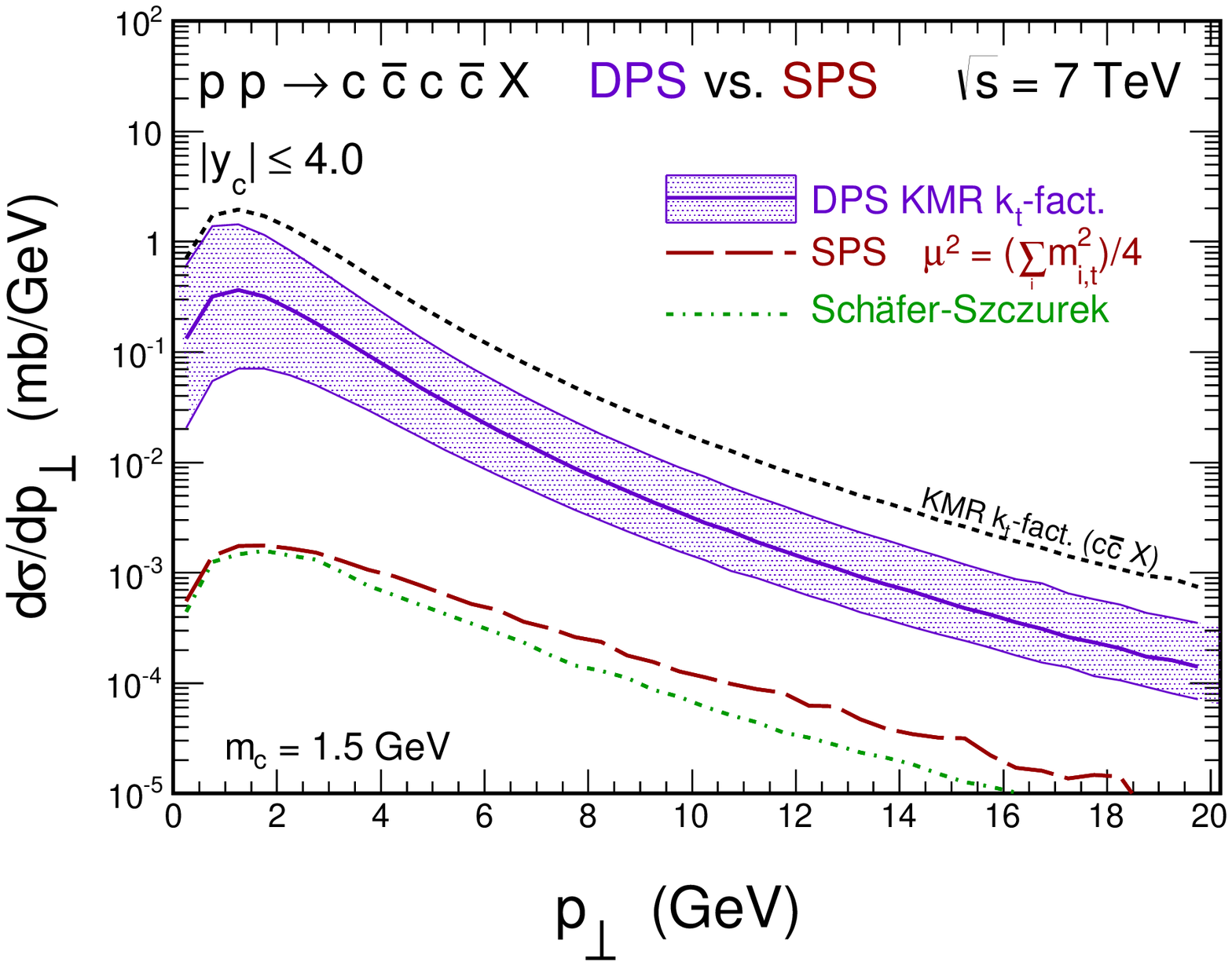}}
\end{minipage}
   \caption{
\small Rapidity (left panel) and transverse momentum (right panel)
distributions of charm quarks for $c \bar c c \bar c$ production.
}
 \label{fig:pt-and-y-spectra}
\end{figure}

In Fig.~\ref{fig:pt-and-y-spectra} we present again rapidity (left panel)
and transverse momentum (right panel) distributions for 
the dominant gluon-gluon $c \bar c c \bar c$ production. 
The uncertainties in DPS contribution are shown as the dashed band. 
The new exact calculation of SPS contribution is shown as the dashed
line and compared to the approximate high-energy Szczurek-Sch\"afer approach
\cite{Schafer:2012tf} shown by the dash-dotted line.
Here in the exact SPS calculations
the factorization and renormalization scales were fixed at
$\mu_R^2 = \mu_F^2 =  (\sum\limits_{i=1}^4 m_{i,\perp}^{2})/4$.
One can clearly see that the SPS contribution is much smaller than 
the DPS contribution,
which confirms that the production of $c \bar c c \bar c$ is an ideal
place to study double parton effects as advocated already in 
\cite{Luszczak:2011zp}.

\begin{figure}[!h]
\begin{minipage}{0.47\textwidth}
 \centerline{\includegraphics[width=1.0\textwidth]{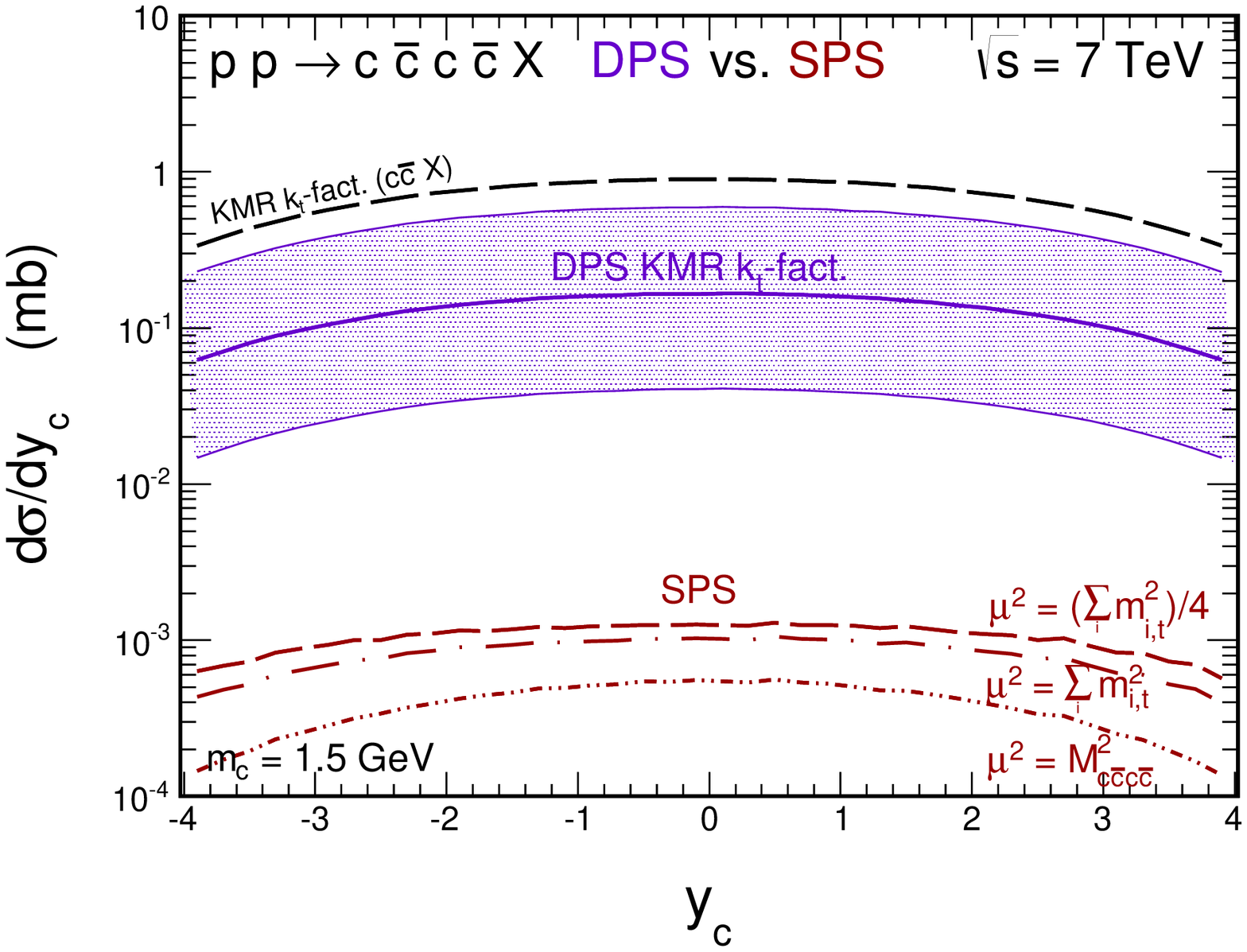}}
\end{minipage}
\hspace{0.5cm}
\begin{minipage}{0.47\textwidth}
 \centerline{\includegraphics[width=1.0\textwidth]{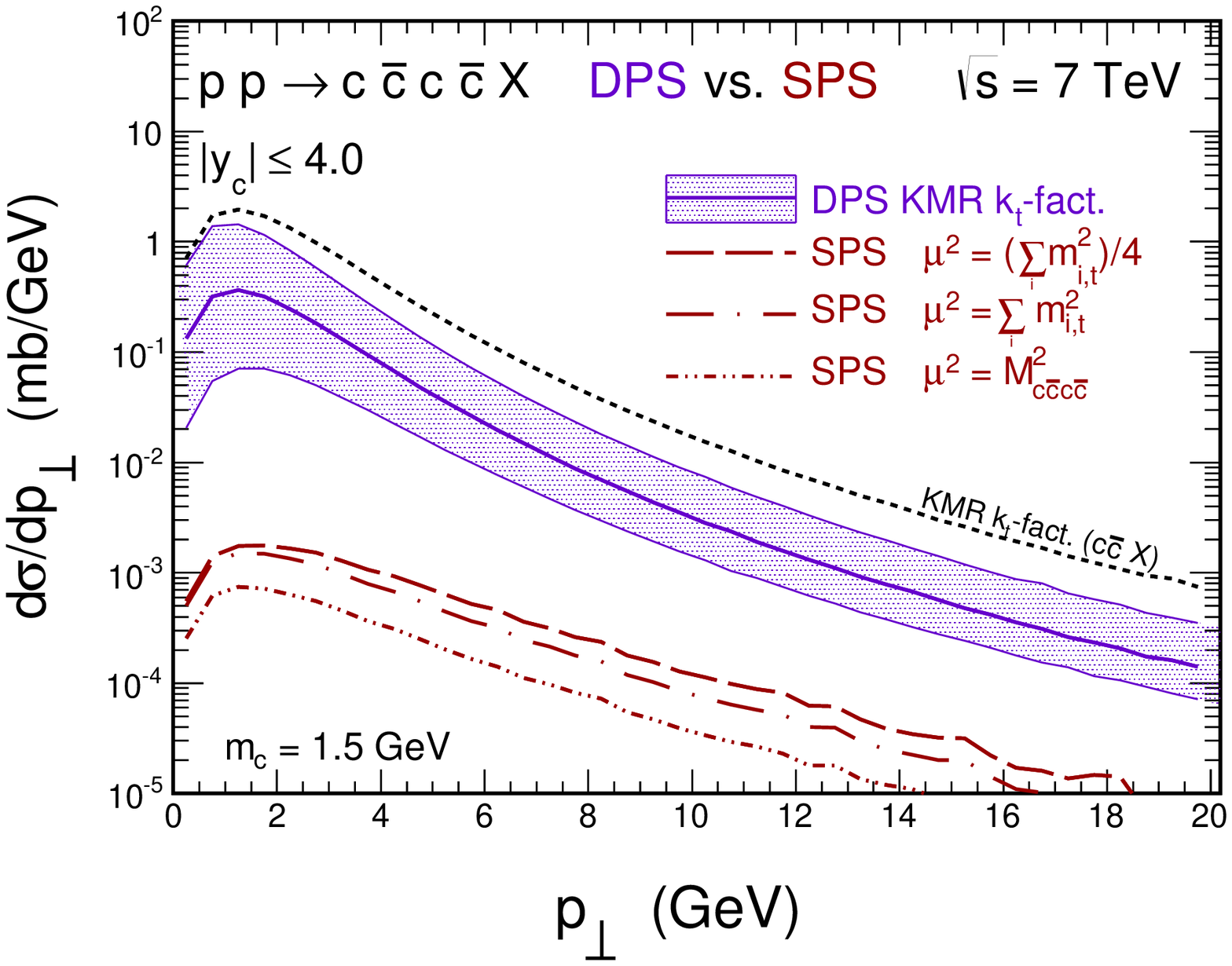}}
\end{minipage}

   \caption{
\small 
Uncertainties of the SPS contribution due to the choice of 
renormalization and factorization scale.
}
 \label{fig:pt-and-y-spectra-scales}
\end{figure}

The dependence on the choice of scales is quantified in 
Fig.~\ref{fig:pt-and-y-spectra-scales}. Uncertainties
up to factor 4 can be observed. The choice 
$\mu_R^2 = \mu_F^2 = M_{c \bar c c \bar c}$ gives the smallest result.

\begin{figure}[!h]
\begin{minipage}{0.47\textwidth}
 \centerline{\includegraphics[width=1.0\textwidth]{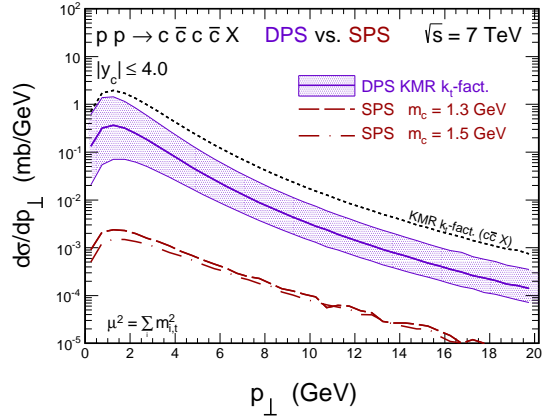}}
\end{minipage}

   \caption{
\small Uncertainties of the SPS contribution due to the choice of
charm quark mass.
}
 \label{fig:pt-spectra-mass}
\end{figure}

In Fig.~\ref{fig:pt-spectra-mass} we present uncertainties on the choice
of charm quark mass. The related uncertainty is much smaller than that
related to the choice of scales.

\begin{figure}[!h]
\begin{minipage}{0.47\textwidth}
 \centerline{\includegraphics[width=1.0\textwidth]{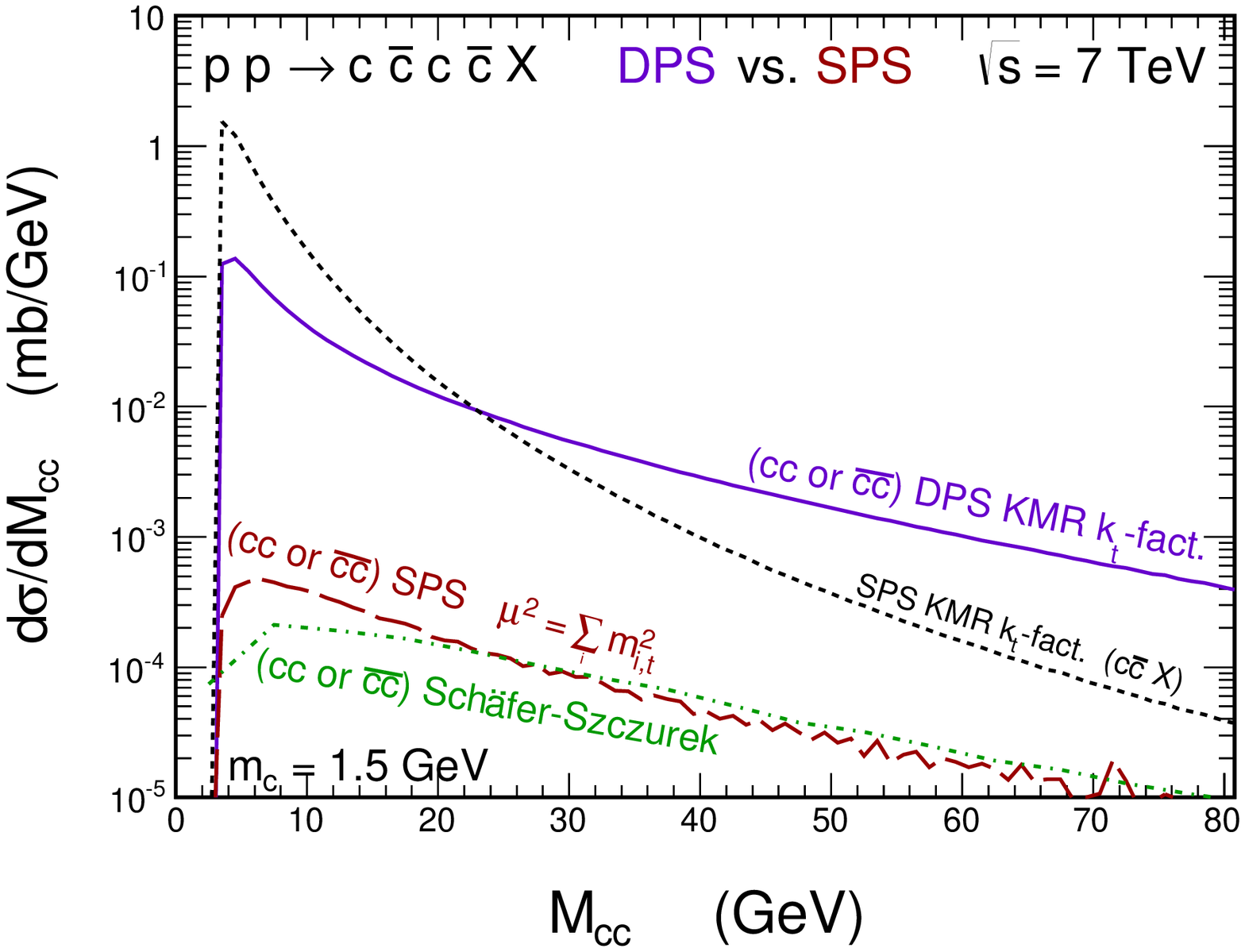}}
\end{minipage}
\hspace{0.5cm}
\begin{minipage}{0.47\textwidth}
 \centerline{\includegraphics[width=1.0\textwidth]{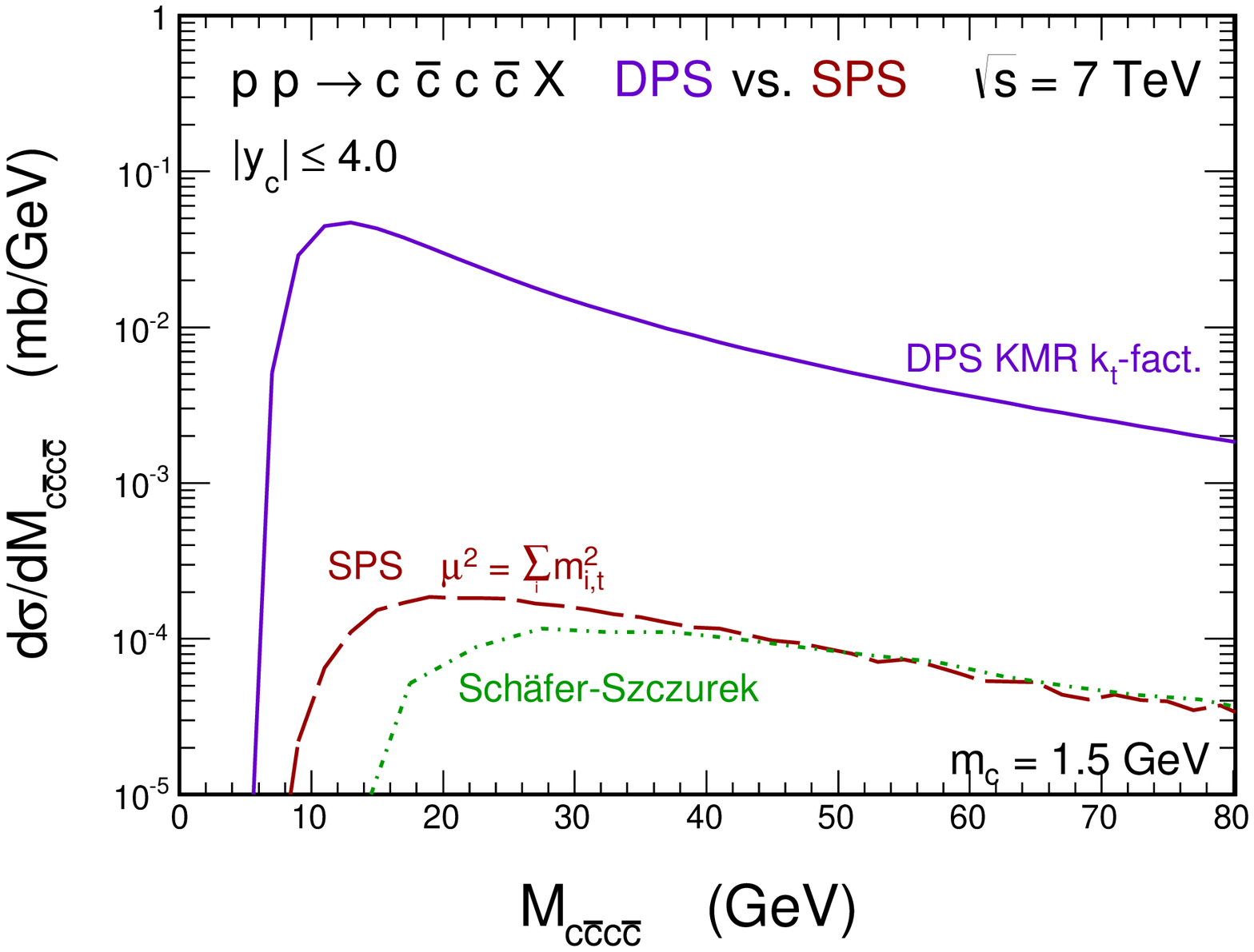}}
\end{minipage}

   \caption{
\small Invariant mass distribution of two $c$ quarks (or two $\bar c$
antiquarks) (left panel) and of the $c \bar c c \bar c$ system 
(right panel) for DPS and SPS with exact and approximate matrix element.
}
 \label{fig:Minv}
\end{figure}

In Fig.~\ref{fig:Minv} we show invariant mass distribution of
the $c c$ (or $\bar c \bar c$) and $c \bar c c \bar c$ systems.
We present separate contributions of DPS and SPS. In addition, 
we compare results of calculations with exact and approximate matrix elements.
One can see a clear difference at small invariant masses,
which is due to absence of the gluon splitting mechanism in the
high-energy approximation of Ref.~\cite{Schafer:2012tf}.

\begin{figure}[!h]
\begin{minipage}{0.47\textwidth}
 \centerline{\includegraphics[width=1.0\textwidth]{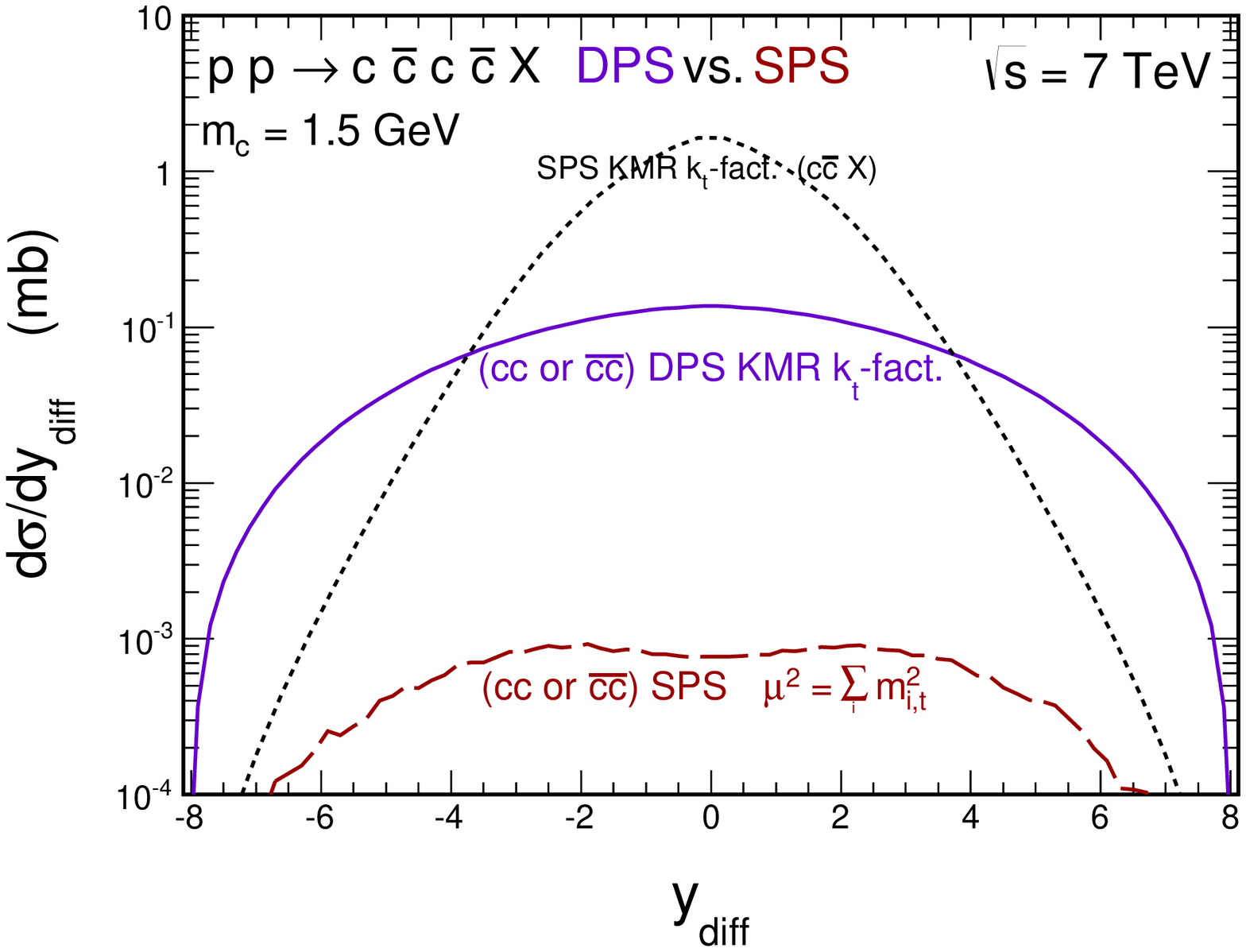}}
\end{minipage}
\hspace{0.5cm}
\begin{minipage}{0.47\textwidth}
 \centerline{\includegraphics[width=1.0\textwidth]{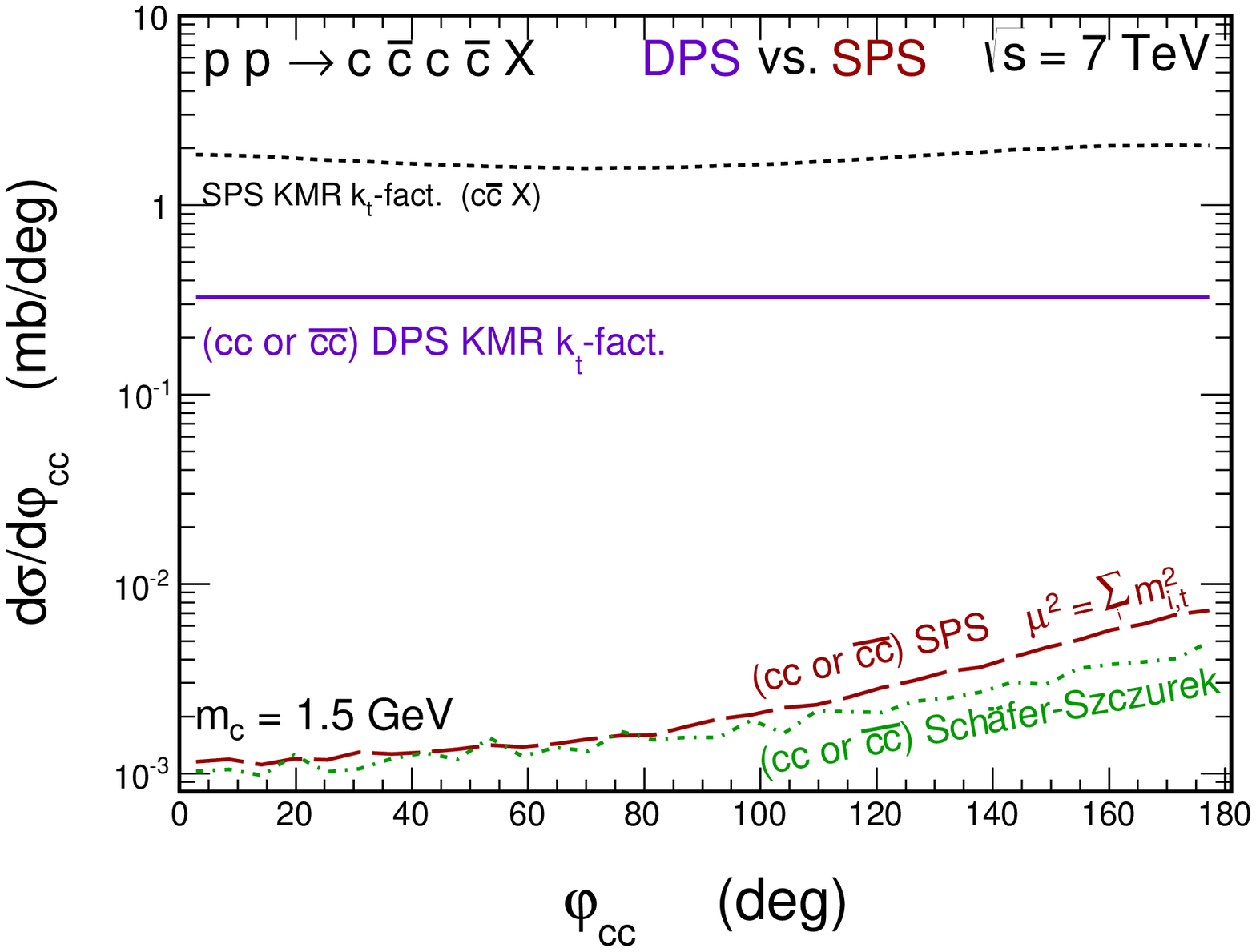}}
\end{minipage}

   \caption{
\small Distribution in the rapidity distance between two $c$ quarks (or two 
$\bar c$ antiquarks) (left panel) and distribution in relative azimuthal
angle between tow $c$ quarks (or two $\bar c$ antiquarks) (right panel).
}
 \label{fig:ydiff-and-phid-spectra}
\end{figure}

Some one-dimensional correlation distributions are shown in 
Fig.~\ref{fig:ydiff-and-phid-spectra}. In the left panel we
present the distribution in rapidity distance between the two $c$ quarks 
(or two antiquarks) and in the right panel the distribution in relative 
azimuthal angle between the two quarks (or two antiquarks).
Both in DPS and SPS large rapidity distances (between $c c$ or 
$\bar c \bar c$) are present. On the other hand, the distributions
in relative azimuthal angle are qualitatively different.
While the DPS contribution is completely flat, the SPS contribution
is rising towards the back-to-back configuration where it takes a maximum.
Since the SPS contribution is much smaller the rise cannot be observed
in real experiments. Both the approximate (Sch\"afer-Szczurek) and the exact
result give very similar dependence in relative azimuthal angle.
For comparison we show also distribution between $c$ and $\bar c$
in SPS production of one $c \bar c$ pair calculated in the
$k_t$-factorization approach with KMR UGDF.

\begin{figure}[!h]
\begin{minipage}{0.47\textwidth}
 \centerline{\includegraphics[width=1.0\textwidth]{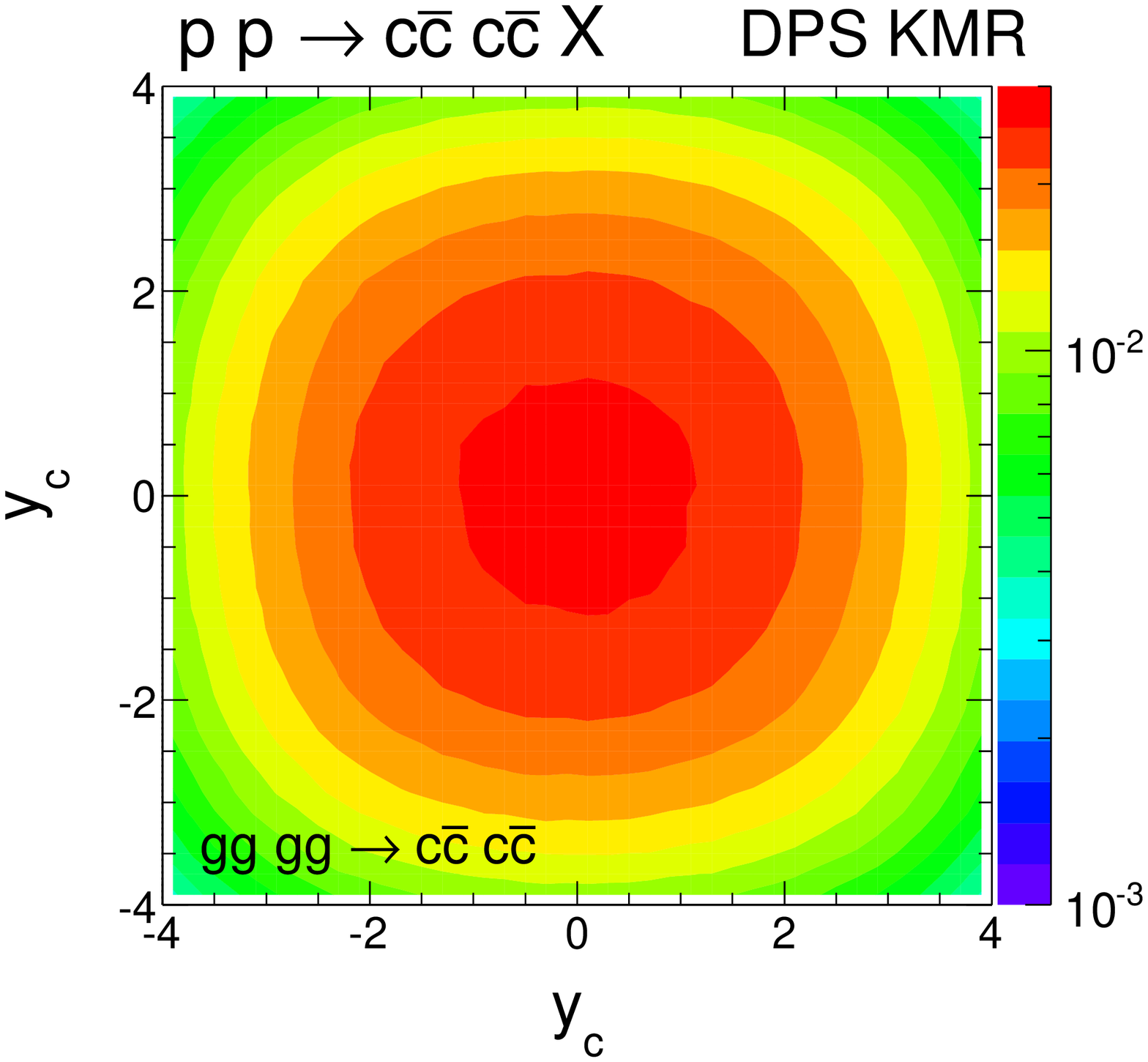}}
\end{minipage}
\hspace{0.5cm}
\begin{minipage}{0.47\textwidth}
 \centerline{\includegraphics[width=1.0\textwidth]{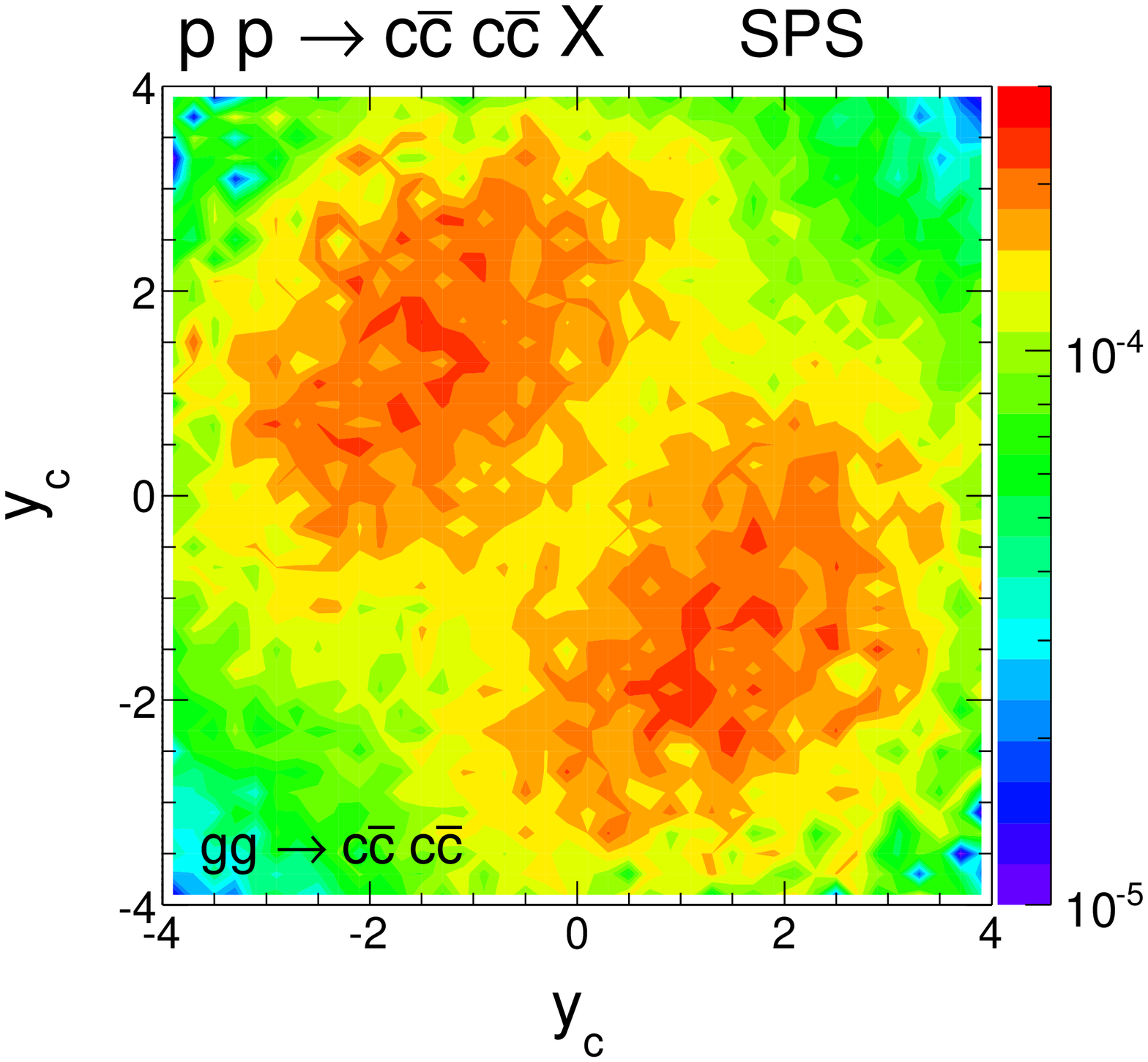}}
\end{minipage}

   \caption{
\small Correlation in rapidities between two $c$ quarks from
the $c \bar c c \bar c$ event for DPS (left panel) and SPS (right panel).
}
 \label{fig:2dim-y1y2-spectra}
\end{figure}

Particularly interesting are correlations in rapidities of two $c$ quarks.
In Fig.~\ref{fig:2dim-y1y2-spectra} we compare results obtained
for DPS (left panel) and SPS (right panel). In our DPS model the two
$c$ quarks are completely decorrelated. In the SPS calculation 
the two quarks are correlated via respective matrix element 
and energy-momentum conservation.

In the next section we shall present results at the hadron level.

\subsection{$D$ meson correlations}

Now we wish to present our results for $D$ mesons.
Below we shall consider only $D^0$ meson production studied 
recently by the LHCb collaboration. 

\begin{figure}[!h]
\begin{minipage}{0.47\textwidth}
 \centerline{\includegraphics[width=1.0\textwidth]{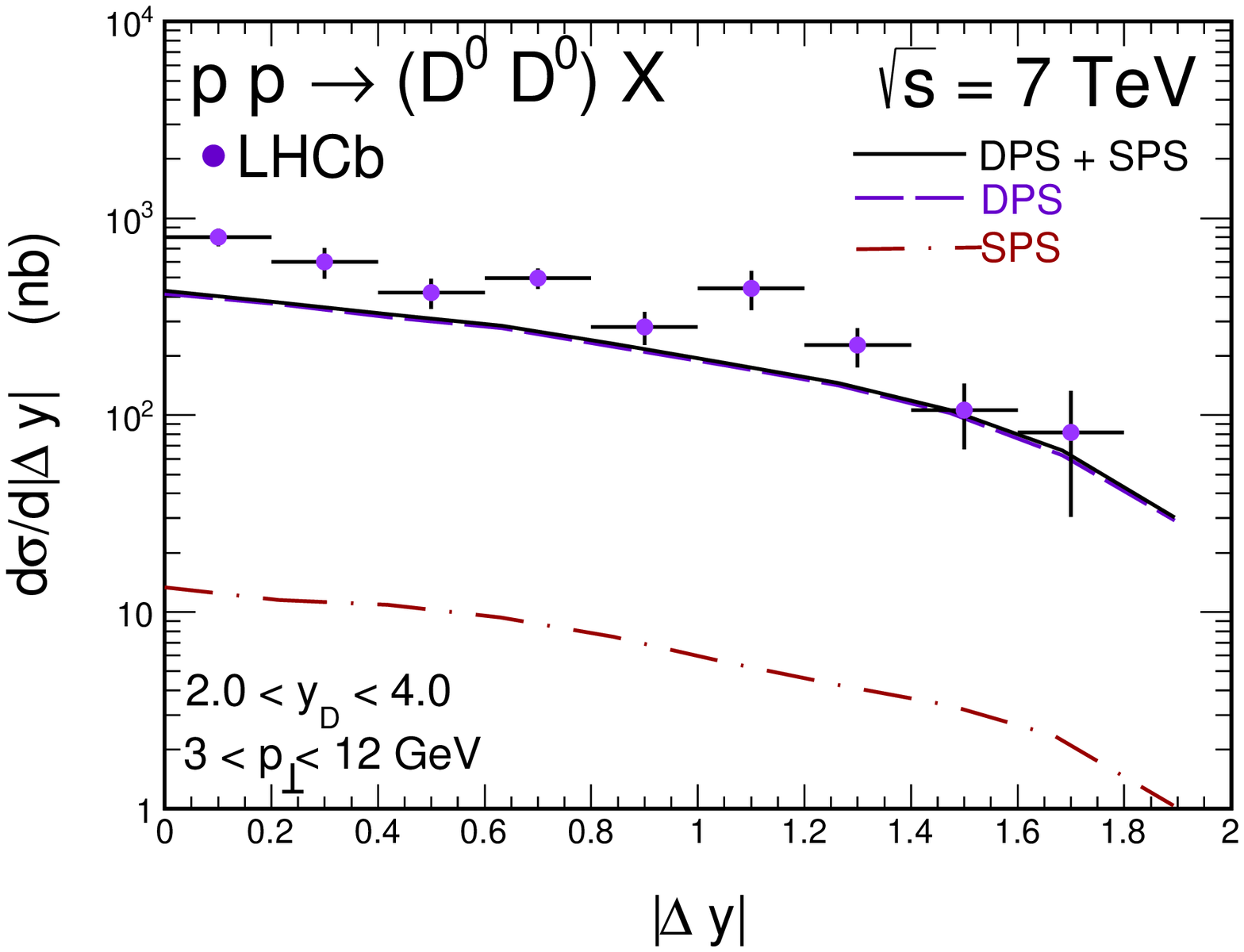}}
\end{minipage}
\hspace{0.5cm}
\begin{minipage}{0.47\textwidth}
 \centerline{\includegraphics[width=1.0\textwidth]{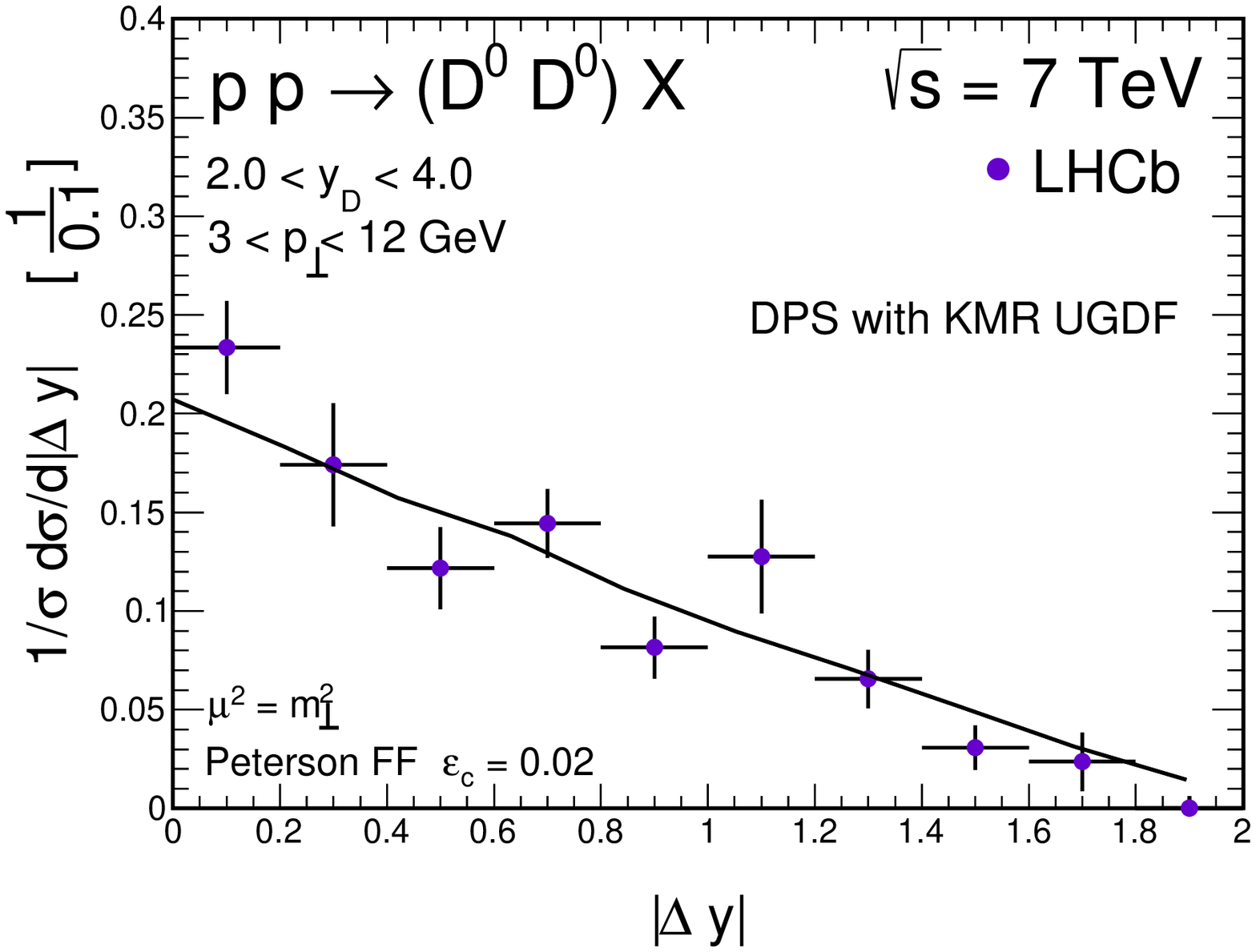}}
\end{minipage}

   \caption{
\small Distribution in rapidity difference between two $D^0$ mesons
(left panel).
The SPS contribution (dash-dotted line) is compared to the DPS
contribution (dashed line).
The right panel shows distribution normalized to the total cross section.
}
 \label{fig:ydiff-mesons}
\end{figure}

In Fig.~\ref{fig:ydiff-mesons} we show distributions in relative
rapidity distance between two $D^0$ mesons with kinematical cuts 
(rapidities and transverse momenta) corresponding to the LHCb
experiment. The distribution normalized to the total cross
section is shown in the right panel. The shape of the distribution is
well reproduced. 

\begin{figure}[!h]
\begin{minipage}{0.47\textwidth}
 \centerline{\includegraphics[width=1.0\textwidth]{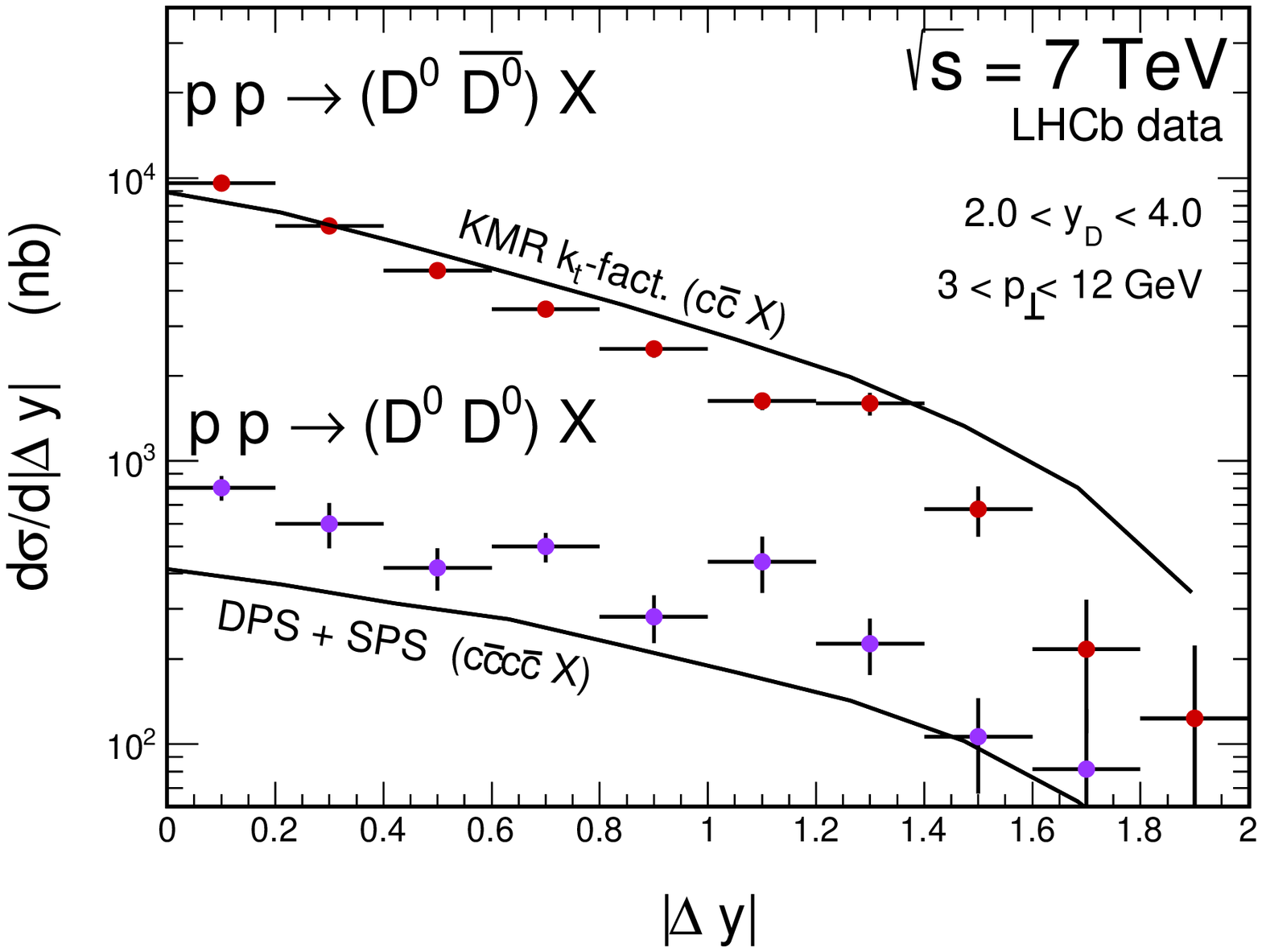}}
\end{minipage}
\hspace{0.5cm}
\begin{minipage}{0.47\textwidth}
 \centerline{\includegraphics[width=1.0\textwidth]{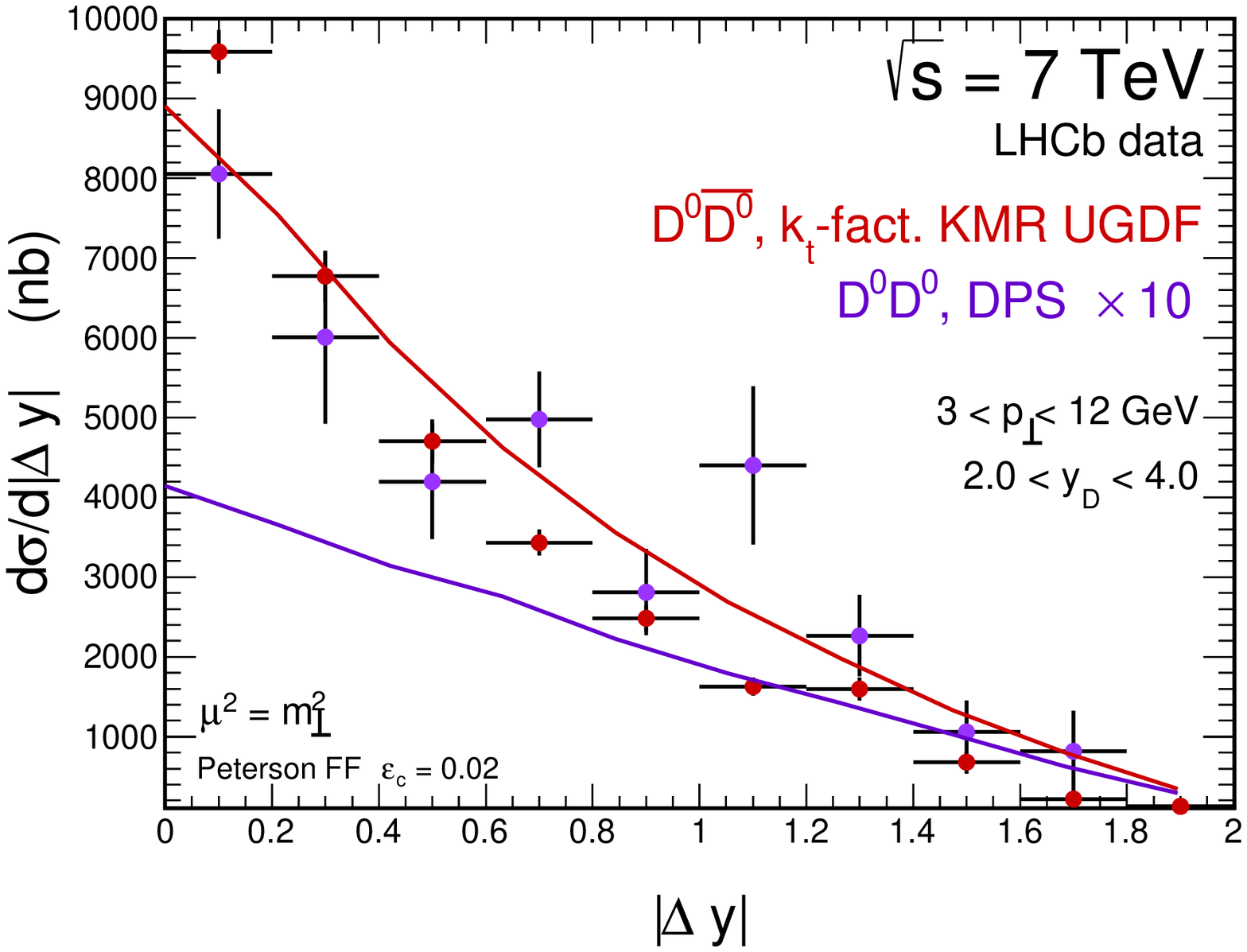}}
\end{minipage}

   \caption{
\small Distribution in rapidity distance between two $D^0$ mesons and
between $D^0 \bar D^0$ from SPS production.
}
 \label{fig:ydiff-mesons-2}
\end{figure}

In Fig.~\ref{fig:ydiff-mesons-2} we compare the distribution in rapidity
distance between $D^0$ and $D^0$ to a similar distribution for the distance
between $D^0$ and $\bar D^0$. The latter distribution falls down
somewhat faster.
This is shown both in logarithmic (left panel) and linear (right panel) scale.

\begin{figure}[!h]
\begin{minipage}{0.47\textwidth}
 \centerline{\includegraphics[width=1.0\textwidth]{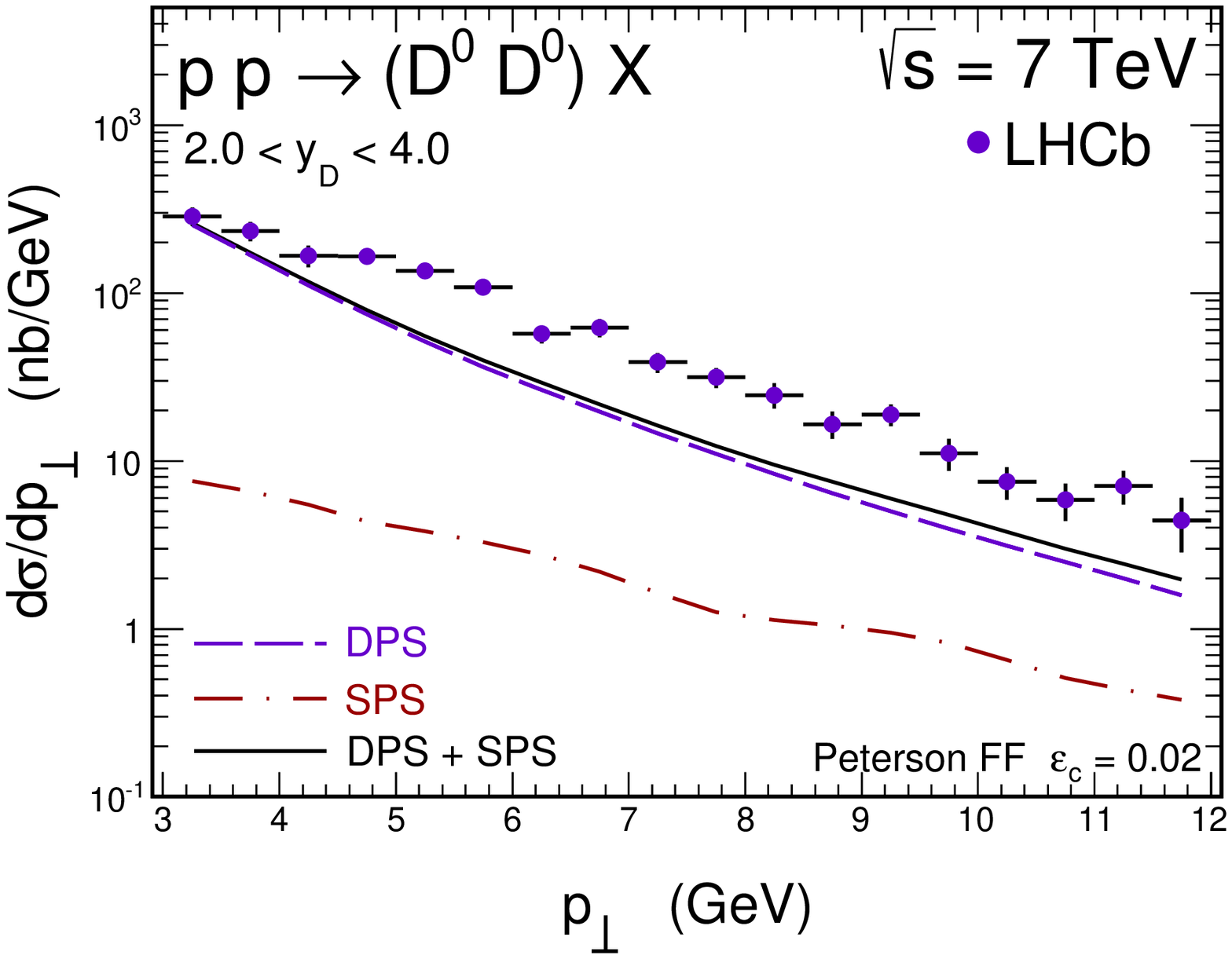}}
\end{minipage}
\hspace{0.5cm}
\begin{minipage}{0.47\textwidth}
 \centerline{\includegraphics[width=1.0\textwidth]{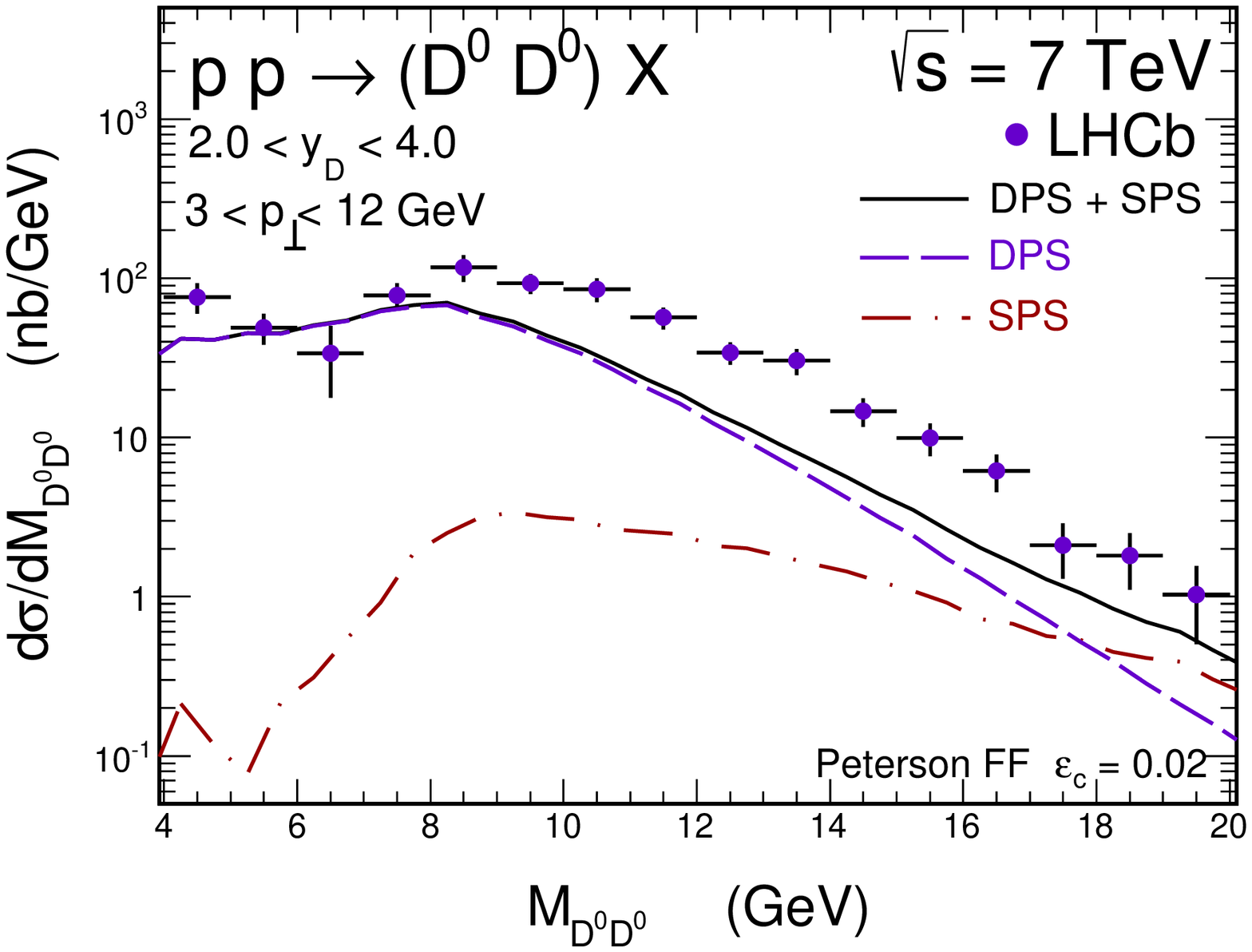}}
\end{minipage}

   \caption{
\small Distributions in meson transverse momentum when both mesons are
measured within the LHCb acceptance and corresponding distribution
in meson invariant mass for DPS and SPS contributions.
}
 \label{fig:Minv-and-pt-mesons}
\end{figure}

Other distributions in meson transverse momentum and two-meson invariant
mass are shown in Fig.~\ref{fig:Minv-and-pt-mesons}.
The shape in the transverse momentum is almost correct but some cross
section is lacking.
Two-meson invariant mass distribution is shown in the right panel.
One can see some lacking strength at large invariant masses.

\begin{figure}[!h]
\begin{minipage}{0.47\textwidth}
 \centerline{\includegraphics[width=1.0\textwidth]{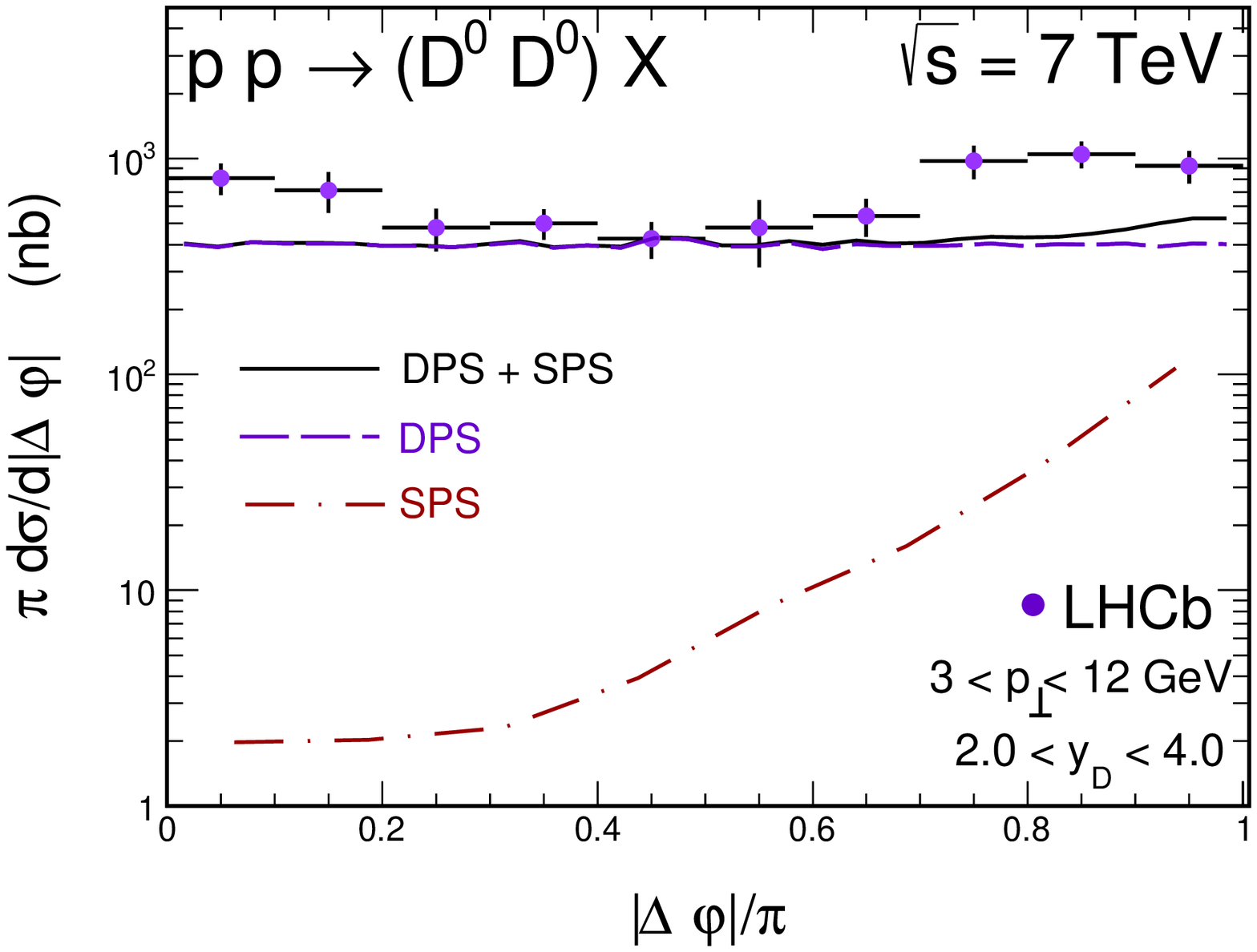}}
\end{minipage}
\hspace{0.5cm}
\begin{minipage}{0.47\textwidth}
 \centerline{\includegraphics[width=1.0\textwidth]{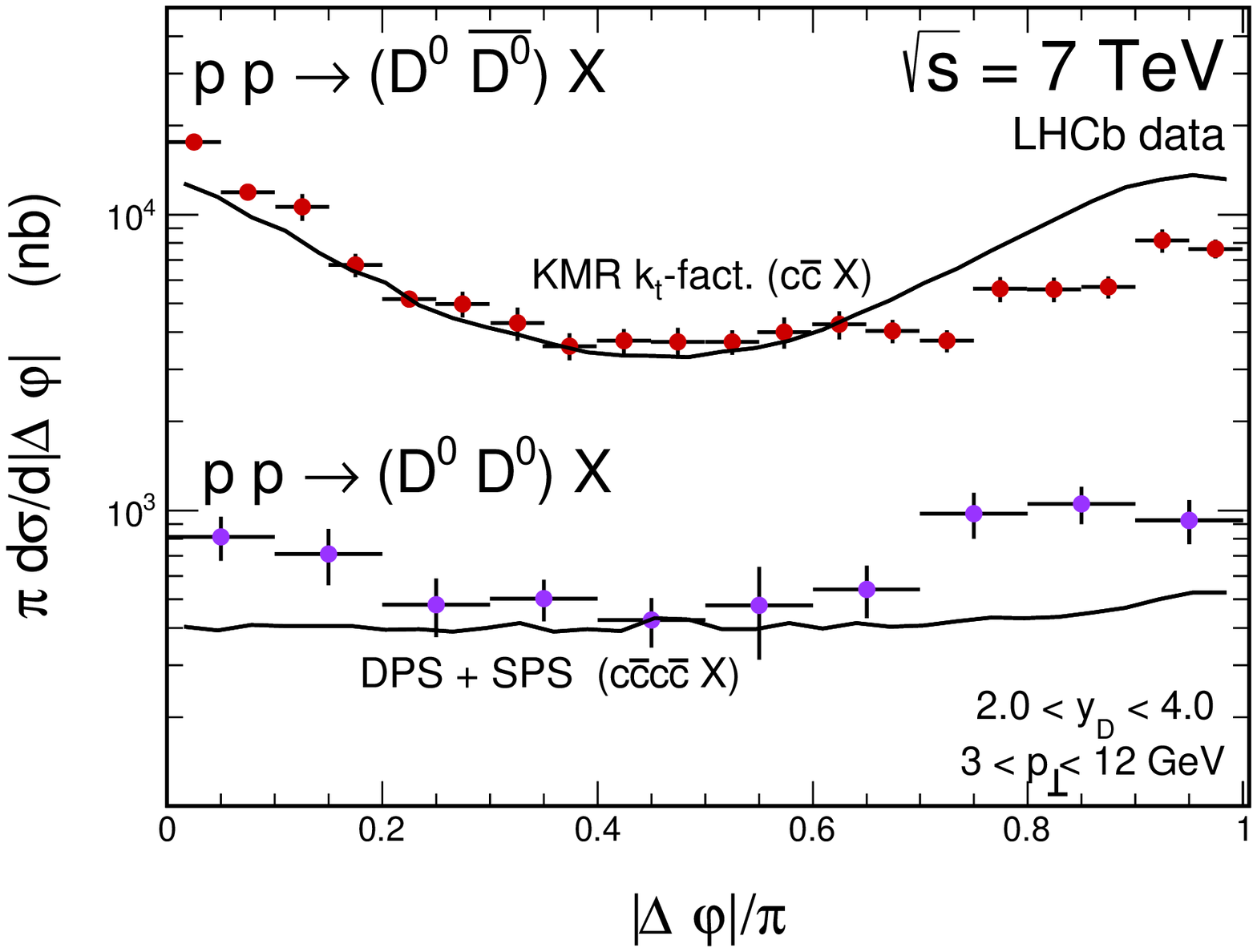}}
\end{minipage}

   \caption{
\small Azimuthal angle correlation between $D^0 D^0$ and $D^0 \bar D^0$.
}
 \label{fig:Phid-mesons}
\end{figure}

We close our presentation with azimuthal angle correlation. In
Fig.~\ref{fig:Phid-mesons} we compare correlations for $D^0 D^0$ and 
$D^0 \bar D^0$. The distribution for identical mesons is somewhat
flatter than that for $D^0 \bar D^0$ which is consistent with the
dominance of the DPS contribution.

\section{Conclusions}

In the present analysis we have compared several distributions
for double charm production for double parton and single parton
scattering. In the latter case we have performed
the calculation with exact matrix element 
for both $g g \to c \bar c c \bar c$ and $q \bar q \to c \bar c c \bar c$
subprocesses (all possible leading-order diagrams).

First, we have discussed results at the quark level. Both one particle
(rapidity, transverse momentum) and correlation
(invariant masses, rapidity distances) distributions have been presented.
We have shown that the results of single parton scattering with
exact and approximate (previously used in the literature) matrix
element differ only in some corners of the phase space, in particular
for small invariant masses of $c c$ and $c \bar c c \bar c$ systems.
The difference can be understood as due to gluon splitting mechanisms 
not present in high-energy approximation \cite{Schafer:2012tf}.
The general situation stays, however, unchanged.
At the LHC energy $\sqrt{s}$ = 7 TeV one observes an unprecedent
dominance of double parton scattering. At the nominal energy $\sqrt{s}$
= 14 TeV the dominance would be even larger.
This opens a possibility to study the DPS effects in the production
of charmed mesons.

In the next stage we have performed a calculation for production
of $D$ mesons. Here we have chosen kinematics relevant for the LHCb
experiment. In contrast to ATLAS or CMS, the LHCb apparatus can measure 
only rather forward mesons but down to very small transverse momenta.
We have calculated several distributions and compared our results
with experimental data of the LHCb collaboration.
Our DPS mechanism gives a reasonable explanation of the measured
distribution. Some strength is still missing. This can be due to
3 $\to$ 4 processes discussed e.g. in Ref.~\cite{3to4-Strikman} 
in the context of four jet production. This will be a subject 
of separate studies.

\vspace{1cm}

{\bf Acknowledgments}

This work was supported in part by the Polish grant
DEC-2011/01/B/ST2/04535 as well as by the Centre for
Innovation and Transfer of Natural Sciences and Engineering Knowledge in
Rzesz{\'o}w.


\end{document}